\documentclass{llncs}
\usepackage[usenames,dvipsnames,dvinames]{xcolor}
\usepackage{amsmath,amsfonts,graphicx,amssymb,sidecap,
color,subfigure,wrapfig,microtype,cancel, times}
\usepackage{cleveref}
\usepackage{algorithm}
\usepackage{algpseudocode}
\usepackage[margin=1in]{geometry}
\usepackage{etoolbox}\AtBeginEnvironment{algorithmic}{\scriptsize}
\algrenewcommand\alglinenumber[1]{\scriptsize #1:}
\usepackage[font=scriptsize,labelfont=bf]{caption}

\usepackage{mdwlist}

\DeclareMathOperator*{\argmax}{argmax}

\newboolean{isdraft}
\setboolean{isdraft}{False}
\ifthenelse{\boolean{isdraft}}{
\newcommand{\john}[1]{\textcolor{red}{(John: #1)}}
\newcommand{\JeTJo}[1]{\textcolor{red}{(Jeff$\to$John: #1)}}
\newcommand{\BTJo}[1]{\textcolor{red}{(Bill$\to$John: #1)}}
\newcommand{\STJo}[1]{\textcolor{red}{(Shengjie$\to$John: #1)}}
\newcommand{\jeff}[1]{\textcolor{green}{Jeff: #1}}
\newcommand{\BTJe}[1]{\textcolor{green}{Bill$\to$Jeff: #1}}
\newcommand{\JoTJe}[1]{\textcolor{green}{John$\to$Jeff: #1}}
\newcommand{\STJe}[1]{\textcolor{green}{Shengjie$\to$Jeff: #1}}
\newcommand{\bill}[1]{\textcolor{blue}{Bill: #1}}
\newcommand{\JeTB}[1]{\textcolor{blue}{Jeff$\to$Bill: #1}}
\newcommand{\JoTB}[1]{\textcolor{blue}{John$\to$Bill: #1}}
\newcommand{\STB}[1]{\textcolor{blue}{Shengjie$\to$Bill: #1}}
\newcommand{\shen}[1]{\textcolor{purple}{(Shengjie: #1)}}
\newcommand{\JeTS}[1]{\textcolor{purple}{(Jeff$\to$Shengjie: #1)}}
\newcommand{\BTS}[1]{\textcolor{purple}{(Bill$\to$Shengjie: #1)}}
\newcommand{\JoTS}[1]{\textcolor{purple}{(John$\to$Shengjie: #1)}}
}{
\newcommand{\john}[1]{}
\newcommand{\JeTJo}[1]{}
\newcommand{\BTJo}[1]{}
\newcommand{\STJo}[1]{}
\newcommand{\jeff}[1]{}
\newcommand{\BTJe}[1]{}
\newcommand{\JoTJe}[1]{}
\newcommand{\STJe}[1]{}
\newcommand{\bill}[1]{}
\newcommand{\JeTB}[1]{}
\newcommand{\JoTB}[1]{}
\newcommand{\STB}[1]{}
\newcommand{\shen}[1]{}
\newcommand{\JeTS}[1]{}
\newcommand{\JoTS}[1]{}
\newcommand{\BTS}[1]{}
}

\providecommand{\doarxiv}{false}
\newboolean{isarxiv}
\setboolean{isarxiv}{\doarxiv} 
\ifthenelse{\boolean{isarxiv}}{%
\newcommand{\arxiv}[1]{#1}
\newcommand{\notarxiv}[1]{}
}{
\newcommand{\arxiv}[1]{}
\newcommand{\notarxiv}[1]{#1}
}
\newcommand{\arxivalt}[2]{\ifthenelse{\boolean{isarxiv}}{#1}{#2}}

\newcommand{\tn}{\tilde{n}}

\newcommand{\expect}{\mathbb{E}}

\newcommand{\pep}{x}
\newcommand{\pepEl}{x}
\newcommand{\candidatePeps}{D}
\newcommand{\pepDb}{\mathcal{D}}

\newcommand{\numero}{n}

\newcommand{\cP}{\mathcal{P}}

\newcommand{\thomson}{\ensuremath{\mathsf{Th}}}
\newcommand{\Nrho}{\widetilde{\numero}^{\pep}}
\newcommand{\obsSpec}{s}

\newcommand{\theoVector}{u}
\newcommand{\theo}{v}
\newcommand{\theoSpec}{\theo^{\pep}}
\newcommand{\procObs}{\tilde{s}}
\newcommand{\obsMz}{m^{\obsSpec}}
\newcommand{\obsCharge}{c^{\obsSpec}}
\newcommand{\NS}{\widetilde{\numero}^{\obsSpec}}

\newcommand{\mzTol}{w}

\newcommand{\Obsmz}{O^{\mbox{mz}}}
\newcommand{\Obsi}{O^{\mbox{in}}}

\newcommand{\ctrain}{\mathcal{C}^i}
\newcommand{\bxt}{s^i}
\newcommand{\bzt}{x^i}
\newcommand{\sSet}{\mathcal{S}}
\newcommand{\xSet}{\mathcal{X}}
\newcommand{\trainSet}{C}

\begin{document}

\mainmatter  

\title{Faster graphical model identification of tandem mass spectra
  using peptide word lattices }

\titlerunning{Lattices for mass spectrum identification}

%
%
\author{Shengjie Wang \and John T.~Halloran \and Jeff A.~Bilmes
\and William S.~Noble}
\authorrunning{Shengjie et al.}

\institute{University of Washington, Seattle, Washington, USA}

%
%

\toctitle{Driptide: Faster Graphical Model Identification}
\tocauthor{Shengjie Wang,John T.~Halloran,Jeff A.~Bilmes,William
  S.~Noble}
\maketitle


\vspace{-0.2in}
\begin{abstract}
Liquid chromatography coupled with tandem mass spectrometry, also
known as {\em shotgun proteomics}, is a widely-used high-throughput
technology for identifying and quantifying proteins in complex
biological samples.  Analysis of the tens of thousands of
fragmentation spectra produced by a typical shotgun proteomics
experiment begins by assigning to each observed spectrum the peptide
that is hypothesized to be responsible for generating the spectrum.
This assignment is typically done by searching each spectrum against a
database of peptides.  We have recently described a machine learning
method---Dynamic Bayesian Network for Rapid Identification of Peptides
(DRIP)---that not only achieves state-of-the-art spectrum
identification performance on a variety of datasets but also provides
a trainable model capable of returning valuable auxiliary information
regarding specific peptide-spectrum matches.  In this work, we present
two significant improvements to DRIP.  First, we describe how to use
{\em word lattices}, which are widely used in natural language
processing, to significantly speed up DRIP's computations.  To our
knowledge, all existing shotgun proteomics search engines compute
independent scores between a given observed spectrum and each possible
candidate peptide from the database.  The key idea of the word lattice
is to represent the set of candidate peptides in a single data
structure, thereby allowing sharing of redundant computations among
the different candidates.  We demonstrate that using lattices in
conjunction with DRIP leads to speedups on the order of tens across
yeast and worm data sets.  Second, we introduce a variant of DRIP
that uses a {\em discriminative training} framework, performing
maximum mutual entropy estimation rather than maximum likelihood
estimation.  This modification improves DRIP's statistical power,
enabling us to increase the number of identified spectrum at a 1\%
false discovery rate on yeast and worm data sets.
\end{abstract}


\vspace{-0.3in}
\section{Introduction}
\vspace{-0.1in}
The most widely used high-throughput technology to identify and
quantify proteins in complex mixtures is {\em shotgun proteomics}, in
which proteins are enzymatically digested, separated by
micro-capillary liquid chromatography and subjected to two rounds of
mass spectrometry.  The primary output of a shotgun proteomics
experiment is a collection of, typically, tens of thousands of {\em
  fragmentation spectra}, each of which ideally corresponds to a
single generating peptide.  The first, and arguably the most
important, task in interpreting such data is to identify the peptide
responsible for generating each observed spectrum.

The most accurate methods to solve this spectrum identification
problem employ a database of peptides derived from the genome of the
organism of interest (reviewed in \cite{nesvizhskii:survey}).  Given
an observed spectrum, peptides in the database are scored, and the
top-scoring peptide is assigned to the spectrum.  We recently proposed
a machine learning method, called DRIP, that solves the spectrum
identification problem using a dynamic Bayesian network (DBN)
\cite{halloran2014uai-drip}.  In this model, the ``time'' axis of the
DBN corresponds to the mass-to-charge (m/z) axis of the observed
spectrum.  The model uses Viterbi decoding to align the observed
spectrum to a theoretical spectrum derived from a given candidate
peptide, while adjusting the corresponding score to account for {\em
  insertions}---a peak in the observed spectrum that is absent from
the theoretical spectrum---and {\em deletions}---a peak in the
theoretical spectrum that is absent from the observed spectrum.  In
DRIP, observed peaks are scored using Gaussians positioned along the
m/z axis, the parameters of which are learned using a training set of
high-confidence peptide-spectrum matches (PSMs).  This approach allows
the model to score observed spectra in their native resolution without
quantization of the m/z axis.

The current work introduces DRIP to the computational biology
community and describes several important improvements to the method.
\begin{itemize}
\item First, and most significantly, we describe how to use {\em word
  lattices} \cite{Ji2006} to make DRIP more efficient.  Word lattices
  are widely used in natural language processing to jointly represent
  a large collection of natural language strings.  In the context of
  shotgun proteomics, a word lattice can be used to represent
  compactly the collection of candidate peptides (i.e., peptides whose
  masses are close to the precursor mass) associated with an observed
  fragmentation spectrum.  Using the lattice during Viterbi decoding
  allows for the sharing of computation relative to the different
  candidate peptides.  Without loss of performance, the lattice
  approach provides a significant reduction in computational expense,
  ranging from 85--93\% in the yeast and worm data sets that we examine
  here.  Notably, this general approach to sharing computations among
  candidate peptides is generally applicable to any scoring function
  that can be framed as a dynamic programming operation.
\item Second, facilitated by the incorporation of word
  lattices, we extend DRIP's learning framework to use discriminative
  training.  Thus, rather than performing maximum likelihood
  estimation on a training set of high-confidence PSMs, DRIP performs
  maximum mutual information estimation to discriminate between the
  high-confidence PSMs and a large ``background'' collection of
  candidate PSMs.  Empirical evidence suggests that this
  discriminative approach provides an improvement in statistical power
  relative to the generatively trained version of DRIP.
\item Third, we introduce several improvements to the model, including
  removing the need to specify several hyperparameters (maximum number
  of allowed insertions and deletions per match) {\em a priori} and a
  calibration procedure to allow joint ranking of PSMs with different
  charge states.
\end{itemize}
The final DRIP model provides an efficient and accurate method for
assigning peptides to observed fragmentation spectra from a shotgun
proteomics experiment, all implemented within a rigorously
probabilistic and easily extensible modeling paradigm.

\vspace{-0.2in}
\section{Overview of Tandem Mass Spectrometry}
\vspace{-0.1in}

\begin{wrapfigure}{l}{4.2in}
\vspace{-30pt}
\begin{center}
\includegraphics[trim=0.0in 0.95in 0.0in 1.5in, clip=true,width=4in]{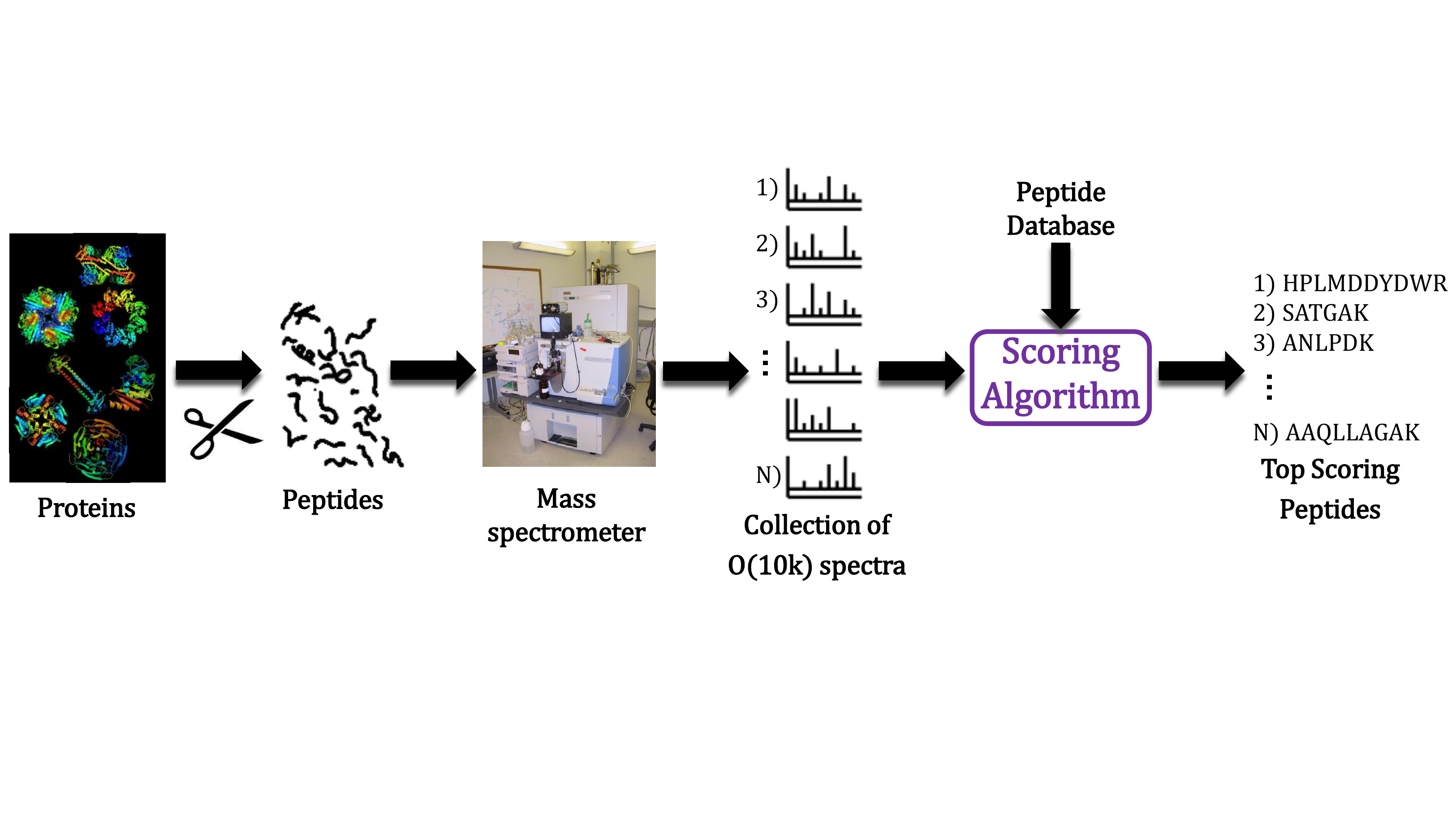}
\end{center}
\vspace{-0.6in}
\caption{A typical shotgun proteomics experiment.}
\label{fig:msmsPipeline}
\vspace{-25pt}
\end{wrapfigure}

A typical shotgun proteomics experiment
(Figure~\ref{fig:msmsPipeline}) begins by cleaving proteins into
peptides using a digesting enzyme, such as trypsin.  The resulting
peptides are then separated via liquid chromatography and injected
into the mass spectrometer.  In the first round of mass spectrometry,
the mass and charge of the intact precursor peptide are measured.
Peptides are then fragmented, and the fragments undergo a second round
of mass spectrometry.  The intensity of each observed peak in the
resulting fragmentation spectrum is roughly proportional to the
abundance of a single fragment ion with a particular m/z value.

Formally, we can represent the spectrum identification problem as
follows.  Let $\cP$ be the set of all possible peptide sequences.
Given an oberved spectrum $\obsSpec$ with precursor mass $\obsMz$ and
precursor charge $\obsCharge$, and given a database of peptides
$\pepDb \subseteq \cP$, we wish to find the peptide $\pep \in \pepDb$
responsible for generating $\obsSpec$.  Using the precursor mass and
charge, we may constrain the set of peptides to be scored by setting a
mass tolerance threshold, $\mzTol$, such that we score the set of
\emph{candidate peptides}
\begin{align}\label{eq:candidateSet}
\candidatePeps(\obsMz, \obsCharge, \pepDb, \mzTol)= \left\{\pep: \pep \in \pepDb ,\, \left| \frac{m(\pep)}{\obsCharge}-\obsMz \right| \leq
\mzTol\right\}, 
\end{align}
where $m(\pep)$ denotes the mass of peptide $x$.  Denoting an
arbitrary scoring function as $\psi(\pep,s)$, the spectrum
identification problem requires finding
\begin{align}
\pep^* = \argmax_{\pep \in  \candidatePeps(\obsMz, \obsCharge,\pepDb,
  \mzTol)} \psi(\pep,s).
\label{eq:spec-id}
\end{align}

\vspace{-0.2in}
\section{Benchmark methods}
\vspace{-0.1in}
To score a peptide $\pep$ with respect to a spectrum $s$, most scoring
algorithms first construct a theoretical spectrum comprised of the
peptide's expected fragment ions.  Denoting the length of $\pep$ as
$n$ and, for convenience, letting $\tn = n-1$, $\pep = x_0 x_1 \dots
x_{\tn}$ is a string.  The fragment ions of $x$ are shifted prefix and
suffix sums called \emph{b-} and \emph{y-ions}, respectively.
Denoting these two as functions $b(\cdot,\cdot,\cdot),
y(\cdot,\cdot,\cdot)$, we have
\begin{align}
b(x,c^b,k) = \mbox{round} \left(\frac{\sum_{i=0}^{k-1}m(\pepEl_i) +
    c^b}{c^b} \right), &&
y(x,c^y,k) = \mbox{round} \left (\frac{\sum_{i=\tilde{n}-k}^{\tilde{n}}m(\pepEl_i)
+18 + c^y}{c^y} \right),
\end{align}
where $c^b$ and $c^y$ are charges related to the precursor charge
state.  The b-ion offset corresponds to the mass of a $c^b$ charged
hydrogen atom, while the y-ion offset corresponds to the mass of a
water molecule plus a $c^y$ charged hydrogen atom.  When $\obsCharge
\in \{1,2\}$, $c^b$ and $c^y$ are unity because, for any b- and y-ion
pair, an ion with no charge is undetectable in the mass spectrometer
and it is unlikely for both charges in a +2 charged precursor ion to
end up on a single fragment ion.  When $\obsCharge \geq 3$, we search
both singly and doubly charged ions so that $c^b, c^y \in \{1,2\}$.
Denoting the number of unique b- and y-ions as $\numero^{\pep}$ and,
for convenience, letting $\Nrho = \numero^{\pep}-1$, our theoretical
spectrum is thus a sorted vector $\theoSpec = (\theo_0, \dots,
\theo_{\Nrho})$ consisting of the unique b- and y-ions of $\pep$.

We benchmark DRIP's performance relative to four widely used search
algorithms.  The score function used by the first such algorithm,
MS-GF+ \cite{kim:generating}, is a scalar product between a heavily
preprocessed observed spectrum, $\procObs$, and a binary theoretical
vector of equal length, $\theoVector$.  MS-GF+ then ranks all the
candidate peptides for a given spectrum by calculating the p-value of
$\theoVector ^T \procObs$ over a distribution of scores for all
peptides with equal precursor mass.
Although the number of distinct peptide sequences with masses close to
the observed precursor mass is large (on the order of
$10^{20}$ in many cases), the linearity of $\theoVector ^T \procObs$
allows the full distribution of scores of all such peptides to be
computed efficiently using dynamic programming.  MS-GF+ version 9980
was used, and the E-value score was used for scoring of spectra.

The remaining three benchmark algorithms are variants of SEQUEST
\cite{eng:approach}, each implemented in Crux v2.1
\cite{mcilwain:crux}.  Like MS-GF+, the SEQUEST algorithm begins by
quantizing and preprocessing the observed spectrum into a vector,
$\procObs$.  A vector of equal length, $\theoVector$, is constructed
based on the theoretical spectrum, and XCorr takes the form
\begin{align}\label{eq:xcorr}
\mbox{XCorr}(\obsSpec, \pep) = \tilde{\theoVector}^T\procObs-
\frac{1}{151}\sum_{\tau=-75}^{75}\tilde{\theoVector}^T\procObs_{\tau}= \tilde{\theoVector}^T(\procObs-
\frac{1}{151}\sum_{\tau=-75}^{75}\procObs_{\tau} )= \tilde{\theoVector}^T\procObs',
\end{align}
where $\procObs_{\tau}$ denotes the vector shifted by $\tau$ m/z
units.  Thus, XCorr is a foreground (scalar
product) minus a background score.  We report results from (1) the
XCorr score as implemented in Tide \cite{diament:faster}, (2) the
XCorr E-value computed by Comet \cite{eng:comet} by fitting, for each
spectrum, an exponential distribution to the candidate peptide XCorr
scores, and (3) the XCorr p-value computed by Tide using a dynamic
programming approach similar to that used by MS-GF+
\cite{howbert:computing}.

\vspace{-0.1in}
\section{Dynamic Bayesian Network for Rapid Identification of Peptides
(DRIP)}
\begin{wrapfigure}{r}{0.5\textwidth}
\vspace{-30pt}
\begin{center}
\includegraphics[trim=0.1in 0.38in 0.05in 0.0in,clip=true,
width=0.99\linewidth]{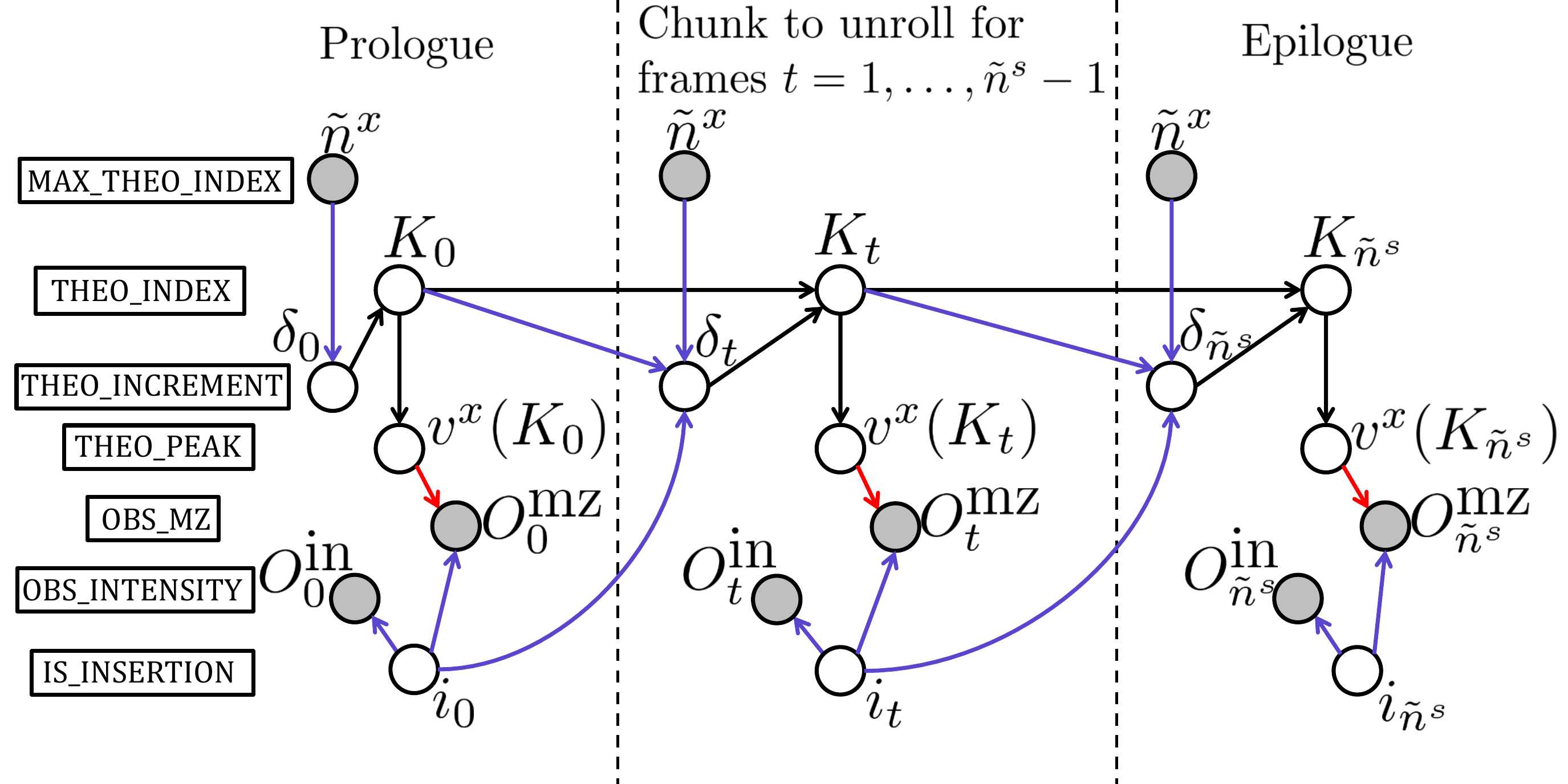}
\end{center}
\vspace{-0.2in}
\caption{Graph of DRIP.}
\label{fig:dripGraph}
\vspace{-20pt}
\end{wrapfigure}

\vspace{-0.1in}
A \emph{graphical model} is a formal representation of the
factorization of the joint distribution governing a set of random
variables via a graph, the nodes of which denote random variables and
the edges of which denote potential conditional dependence between
nodes \BTJo{\cite{FIXME}}.  This formalization enables a host of
tractable inference algorithms while offering incredible modeling
flexibility.  A \emph{Bayesian network} is a graphical model defined
over directed acyclic graphs, and a \emph{dynamic Bayesian network}
(DBN) is a Bayeian network defined over variable length temporal
sequences.  The basic time unit of a DBN, determined by the time units
of the temporal process being modeled, is called a \emph{frame} and
consists of a group of nodes and edges amongst these nodes.  A DBN is
often defined in terms of a \emph{template}, where the first and last
frame are called the \emph{prologue} and emph{epiloque}, respectively,
and where the \emph{chunk} is expanded to occupy all frames in
between.  The template of DRIP is depicted in
Figure~\ref{fig:dripGraph}.

The DRIP model itself represents the observed spectrum, and each frame
in DRIP corresponds to a single observed peak.  The theoretical
spectrum is hidden and traversed from left to right as follows.
Denoting the number of frames as $\numero^{\obsSpec}$ and, for
convenience, letting $\NS=\numero^{\obsSpec}-1$ and $t$ be an
aribtrary frame $0 \leq t \leq \NS$, $K_t$ denotes the index of the
current theoretical peak, which is a deterministic function of its
parents such that $K_0 = \delta_0$ and $K_t = K_{t-1}+\delta_t$ for
$t>0$, where $\delta_t$ is a multinomial random variable.  Thus,
$\delta_t$ denotes the number of theoretical peaks we traverse in
subsequent frames.  Furthermore, the parents of $\delta_t$, $\Nrho$
and $K_{t-1}$, constrain it from being larger than the number of
remaining theoretical peaks left to traverse.  $\theoSpec(K_{t})$ is
thus the $K_t$th theoretical peak.  The variables $\Obsmz_t$ and
$\Obsi_t$ are the observed m/z and intensity values, respectively, of
the $t$th observed peak, where $\Obsmz_t$ is scored using a Gaussian
centered near $\theoSpec(K_{t})$ and $\Obsi_t$ is scored using a
Gaussian whose learned variance is larger than that of the m/z
Gaussians, thus prioritizing matching well along the m/z axis as
opposed to high-intensity peaks.  Parent to these observed variables
is $i_t$, a Bernoulli random variable which denotes whether an
observed peak is considered an insertion, $i_t=1$, or not.  When $i_t
= 1$, a constant penalty is returned rather than scoring the observed
observations with the currently considered Gaussians dictated by
$K_t$, since scores may become arbitrarily bad by allowing Gaussians
to score m/z observations far from their means and this would make the
comparability of scores impossible (a single insertion would make an
otherwise great alignment terrible).  Furthermore, so as to enforce
that some observations be scored rather than inserted, $i_t$ enforces
its child $\delta_{t+1}$ to be zero in a frame following an insertion.
Because $\delta_t$ and $i_t$ are hidden, DRIP considers all possible
{\em alignments}, i.e., all possible combinations of insertions and
theoretical peaks scoring observed peaks.  The Viterbi path, i.e., the
alignment which maximizes the log-likelihood of all the random
variables, is used to score a peptide as well determine which observed
peaks are most likely insertions and which theoretical peaks are most
likely deletions.

Note that the version of DRIP used in this work
(Figure~\ref{fig:dripGraph}) is somewhat simplified relative to the
previously described DRIP model \cite{halloran2014uai-drip}.  In
particular, the previously described version of DRIP required two
user-specified quantities, corresponding to the maximum allowable
numbers of insertions and deletions.  These constraints were enforced
via two chains of variables, which kept track of the number of
utilized insertions and deletions in an alignment, counting down in
subsequent frames.  The current, simplified model automatically
determines these two quantities on the basis of the deletion and
insertion probabilities as well as the insertion penalties.  This
simplification comes at no detriment to performance or running time
(data not shown).

\vspace{-0.1in}
\subsection{Calibrating DRIP with respect to charge}
For observed fragmentation spectra with low-resolution precursor data
often have indeterminate charge states.  For such spectra, all
candidates are scored and ranked assuming all charge states, with the
highest scoring peptide amongst all charges returned.  This approach
requires that scores among differently charged spectra be comparable
to one another.  In DRIP, however, this is not the case.  Because the
number of theoretical peaks essentially doubles when moving from
charge 2+ to charge 3+, higher charged PSMs which have denser
theoretical spectra and thus much better scores than lower charged
PSMs, rendering differently charged PSMs incomparable.

In order to alleviate this problem, we recalibrate scores on a per
charge basis, projecting differently charged PSM scores to roughly the
same range.  The procedure consists of first generating a secondary
set of decoy peptides disjoint from the target peptide set and primary
decoy peptide sets.  Let $\pepDb^t$ be the set of target peptides,
$\pepDb^d$ be the set of decoy peptides, and $\pepDb^{dd}$ be our new
set of decoy peptides, such that $\pepDb^{dd} \cap ( \pepDb^{t} \cup
\pepDb^{d} ) = \emptyset$.  Next, all spectra and charge states are
searched, returning top ranking sets of PSMs $\mathcal{X}^t \subseteq
\pepDb^{t}$, $\mathcal{X}^d \subseteq \pepDb^{d}$, and
$\mathcal{X}^{dd} \subseteq \pepDb^{dd}$.  We then partition
$\mathcal{X}^{dd}$ based on charge.  For each charge $c$ and
corresponding partition, $\mathcal{Z}^{dd}$, we rank all PSMs based on
score.  Let $\mathcal{Z}^{t} \subseteq \mathcal{X}^t$ be all PSMs of
charge $c$, for which we use their scores to linearly interpolate
their normalized ranks in $\mathcal{Z}^{dd}$, using these interpolated
normalized ranks as their new calibrated scores.  In order to
recalibrate scores greater than those found in $\mathcal{Z}^{dd}$, we
use the 99.99th percentile score and maximal score for linear
interpolation.  Once this procedure is done, a majority of
recalibrated scores for PSMs in $\mathcal{Z}^{t}$ and
$\mathcal{Z}^{d}$ will lie between $[0,1]$ and a few barely above
unity(this is largely true for the targets, since a set of well
expressed and identifiable target peptides typically outscores all
decoys), thus greatly decreasing the dynamic range of scores.  With
these recalibrated scores, we may easily take the top ranking PSM of
an observed spectrum amongst differently charged PSMs, without any
loss in overall accuracy.

\vspace{-0.15in}
\section{Word Lattices}
\vspace{-0.1in}
\emph{Word lattices} (abbreviated as lattices for this section)
 represent data instances compactly in graphical structures
\BTJo{This is a pretty vague definition.}  and have gained great
success in the fields of natural language processing and speech
recognition.  For example, natural language dictionaries can be stored
in lattices for more efficient querying; in speech recognition, 
lattices constructed out of the top phone/word level hypotheses
can be used to rescore and select the best hypothesis effectively.
 \BTJo{I don't actually understand the previous sentence.
  How does storing some hypotheses in a lattice help you to get better
  scores?}

A lattice over an alphabet $\Sigma$ is a directed graph $G = (N, E, s,
t)$, where $N$ is the node set, $E$ is the edge set, and $s, t\in N$
denote the source and target node respectively. \BTJo{The notion of
  ``nodes'' and ``edges'' have not yet been introduced.}  Every edge
$e \in E$ is a tuple $(n_1, n_2, \alpha(e))$, where $n_1, n_2$ are the
from-node and to-node respectively, and $\alpha(e) \in \Sigma$ is the
alphabet element encoded in $e$.

A lattice encodes data on paths from $s$ to $t$. Let $P(G, s, t)$
denote the set of paths from $s$ to $t$, then every $p =
e_{1},e_{2},..,e_{|p|}$, $p \in P(G, s, t)$ represents a sequence of
characters, or a string, over alphabet $\Sigma$:
$\alpha(e_1),\alpha(e_2),..,\alpha(e_{|p|})$. For notation simplicity,
let $P(G, e_1, e_2)$, where $e_1, e_2 \ in E$, denote the set of paths
starting in $e_1$ and ending in $e_2$.

An alternative way to define a lattice is to consider it as a
nondeterministic finite automaton (NFA) $G = (Q, \Sigma, \Delta, q_0,
F)$, where $Q$ denotes the set of states, which is equivalent to the
node set in the directed graph definition, $\Delta$ is the transition
function: $Q \times \Sigma \rightarrow 2^Q$, $q_0$ is the initial
state and $F = \{f\}$ is the final state. $q_0$ and $f$ correspond to
$s,t$ in the directed graph definition respectively. The NFA
definition is useful when we consider the problem of constructing a
lattice from a set of strings.


\begin{figure}
  \begin{minipage}[b]{0.55\linewidth}

\centering
\includegraphics[page=1,trim=0.1in 0.1in 0.1in 0.2in,clip=true,
width=0.4\linewidth]{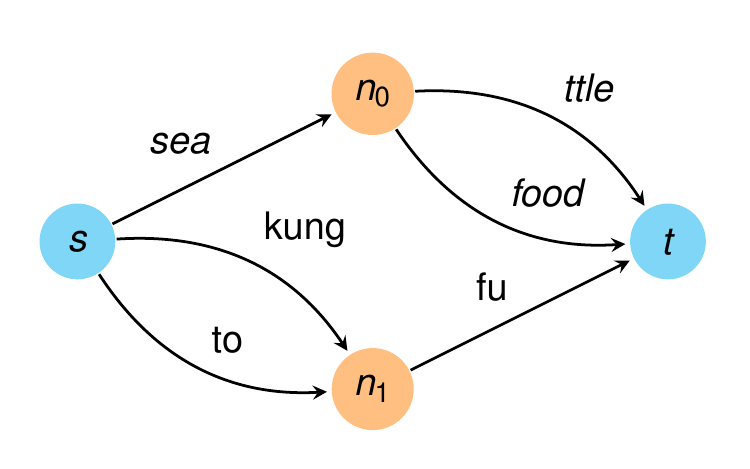}
\caption{An Example Lattice: encodes ``seattle'', ``'seafood',
  ``kungfu'', and ``tofu''; the alphabet can be any string over [a-z]
  of length $\leq 4$.}
\label{fig:latExample}

  \end{minipage}%
  \begin{minipage}[b]{0.35\linewidth}

\hspace{0.25in}
\includegraphics[page=1,trim=0.0in 0.1in 0.0in 0.2in,clip=true,
width=0.7\linewidth]{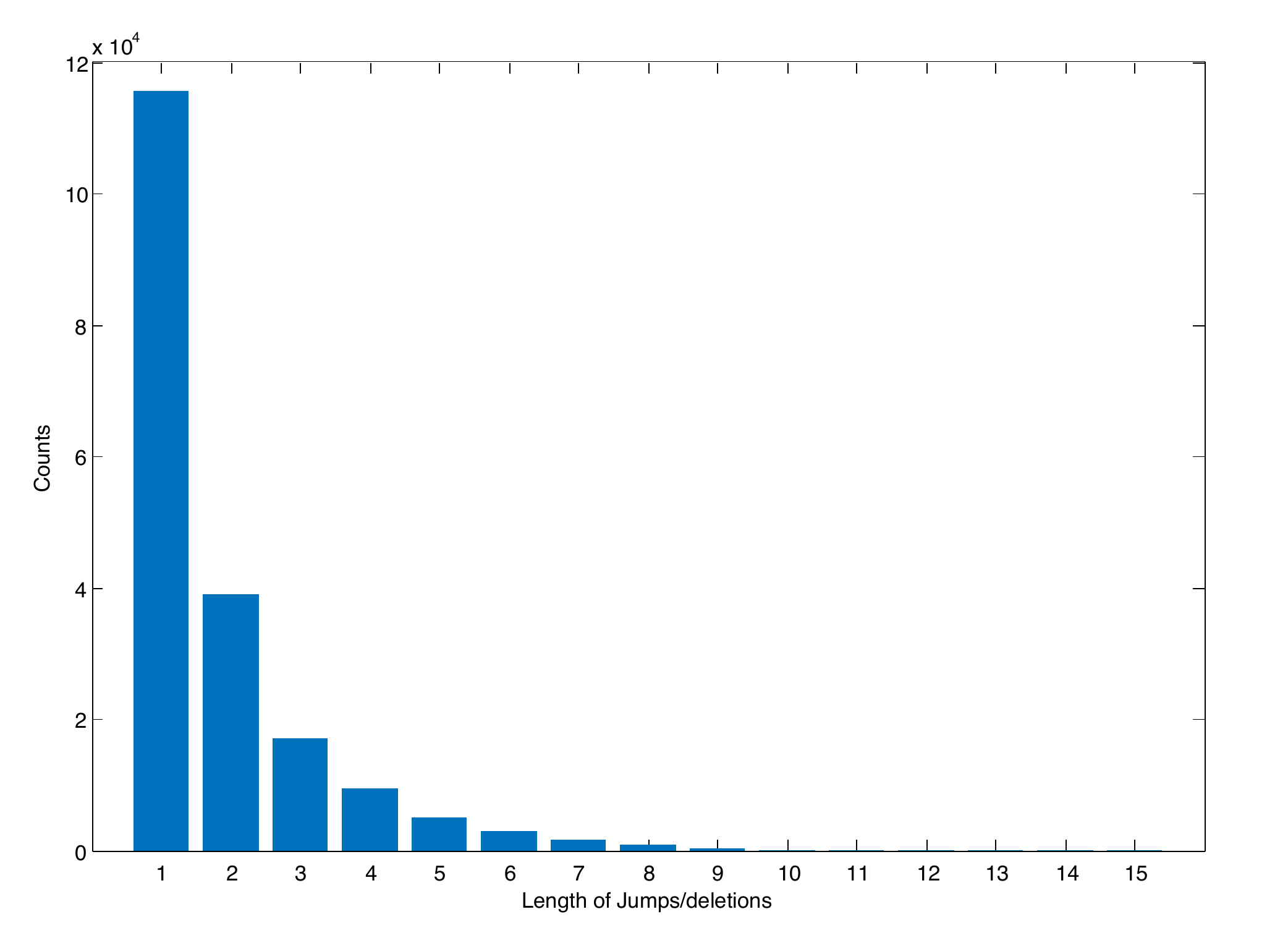}
\vspace{-0.1in}
\caption{Jump Histogram}
\label{fig:jump}

\end{minipage}
\vspace{-0.25in}
\end{figure}

Suppose we are given a set of strings $X = \{x_1, x_2, ...,
x_m\}$. The lattice representation of the strings $G(X)$ can be
extremely compressed as there may be significant amount of redundant
information shared among elements of $X$. More specifically, if $x_i$
and $x_j$ share some substring in common, we can merge the shared
parts into a sequence of common edges, instead of representing the
same substring twice. Fig.~\ref{fig:latExample} gives a lattice over four word strings, with the shared substrings collapsed into common edges to reduce the redundancy.

The common edges of lattices not only save space for data
representation, but also speed-up complex operations using the
encoded data. For this paper in particular, we focus on the task of dynamic graphical
model inference with Viterbi algorithm. With lattices, we manage to
both reduce the state space of Viterbi algorithm, and apply smart
pruning strategies in a more effective way, as we will discuss in
details later, which result in magnitudes of speed-ups.

\vspace{-0.1in}
\subsection{Lattice Construction}

To construct the optimal lattice from input set of strings $X$
over alphabet $\Sigma$ is a hard problem. The objective of the
``optimal'' lattice is task dependent. E.g., for dictionary queries,
the optimal lattice should have the minimal number of nodes/states,
which correspond to the major computation time; for data compression,
the optimal lattice should be minimal in overall size, in which case both
the nodes and edges matter. Moreover, some tasks do not require the
lattice to be an ``exact'' representation of the input strings. For a
set of strings $X$, an \emph{exact lattice} is an NFA that accepts
only language $\mathbb{L}(G(x)) = X$. For our task, which is to
speed-up graphical model training/inference, we choose the objective
to be: construct lattice $G(X) = (N, E, s, t)$ such that $G(X)$ is
exact and $|E|$ is minimized.

As stated above, we can think a lattice as an NFA, and it has been
proven that NFA minimization in terms of number of states/transitions
are NP-hard to approximate within constant factors. To approximate the
optimal lattice, we give the following algorithm which is similar to
the determinize-minimize algorithm of minimizing a DFA\cite{watson1993b}. The
result lattice $G(X)$ is the minimum state DFA of the same language
$X$.

\vspace{-0.2in}
\begin{algorithm}
\caption{$ConstructLattice(X, \Sigma)$}
\label{lattice_con}
\begin{algorithmic}[1]
	\State $G = \{\{s,t\}, \emptyset, s, t\}$.
	\State Assign an ordering in $\Sigma$, and sort $X$.
	\For {$x \in X$}
		\State Start from $s$, traverse $G$ with $x$. Stop
                when no matching edge can be found: suppose the
                non-matching character is $x[i]$, and we stop at node
                $n'$.
		\State Build a chain structure $C$ out of $x[i:]$,
                where each edge on the chain corresponds to one
                character.
		\State Merge $C$ into $G$ by merging the start node of
                $C$ with $n'$ and the end node of $C$ with $t$.
	\EndFor

	\State Run DFA minimization on $G$.\\
	\Return $G$.
\end{algorithmic}
\end{algorithm}
\vspace{-0.2in}

The for loop, which merges prefixes of input strings, constructs a DFA
out of $X$. Minimization on the constructed DFA can be thought as a
process that merges nodes, which share the same suffixes. Both merging
prefixes and suffixes reduce the number of edges in the lattice,
making the algorithm a powerful heuristic in practice, and managing to reduce
up to 50\% edges as we will show in the result section.
The complexity of the algorithm is bounded by the DFA minimization
step. With Hopcroft's algorithm \cite{JohnHopcroft1971}, the running time is
$\mathcal{O}(|\Sigma||X|l_{\max}log(|X|l_{\max}))$, where $l_{\max} =
\max_{x \in X} |x|$.

\vspace{-0.1in}
\subsection{Representation of Lattices in Dynamic Graphical Models}
We can utilize dynamic graphical model structures to naturally traverse a
lattice $G$ to access the encoded data. Particularly, for time frame
$t$, we use three vertices to access the lattice: the lattice node
vertex $V_t$, the lattice link vertex $L_t$, and the transition vertex
$T_t$. $V_t$'s and $T_t$'s decide the set of values that $L_t$'s can
take on, and each value of $L_t$ corresponds to one character in the
encoded strings.



\begin{figure}
  \begin{minipage}[b]{0.50\linewidth}

\centering
\includegraphics[page=1,trim=0.0in 0.0in 0.0in 0.0in,clip=true,
width=0.7\linewidth]{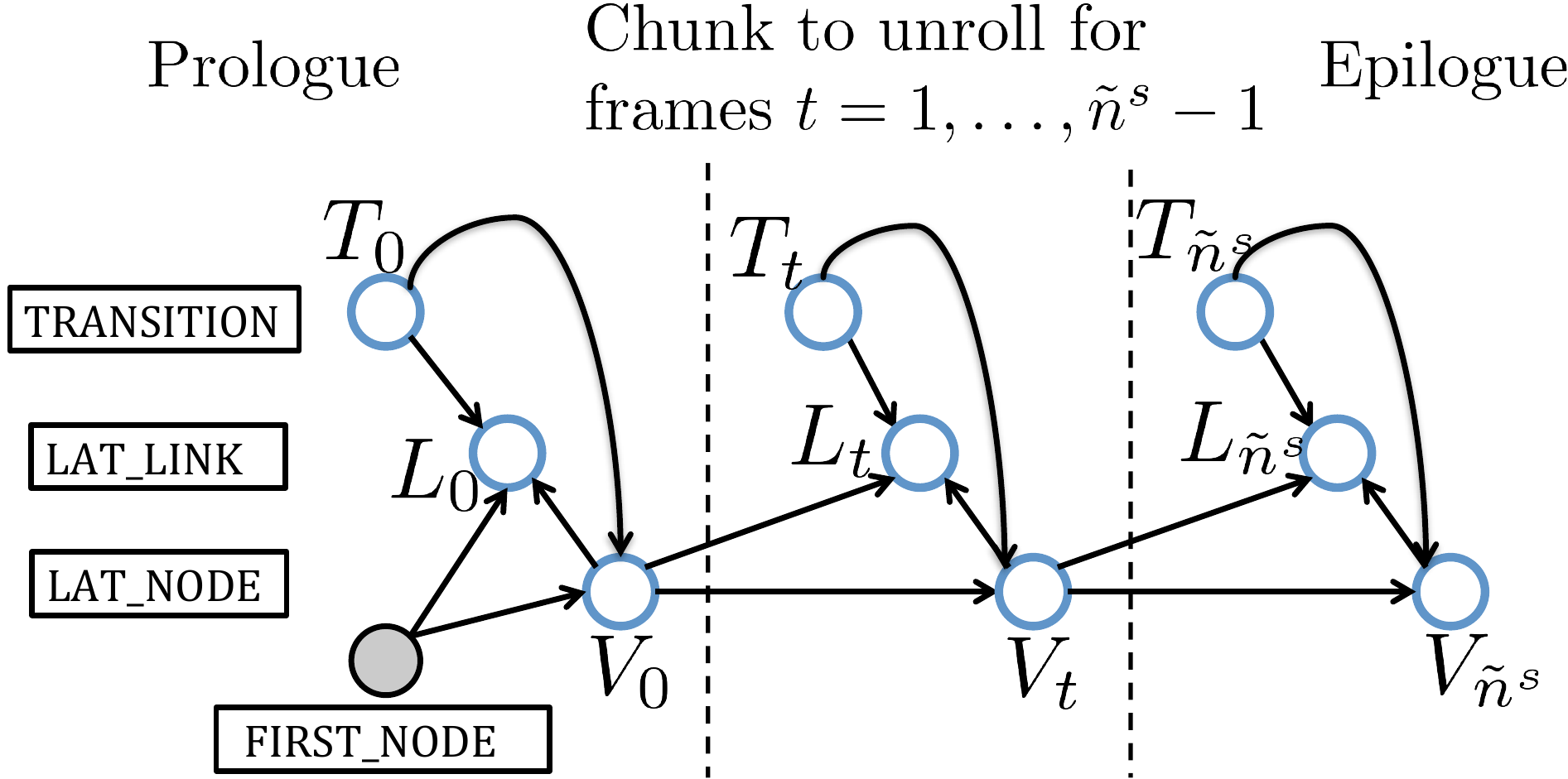}
\vspace{-0.1in}
\caption{Representation of Lattices in Dynamic Graphical Models}
\label{fig:latRep}

  \end{minipage}%
  \begin{minipage}[b]{0.50\linewidth}

\includegraphics[page=1,trim=0.0in 0.25in 0.0in 0in,clip=true,
width=0.75\linewidth]{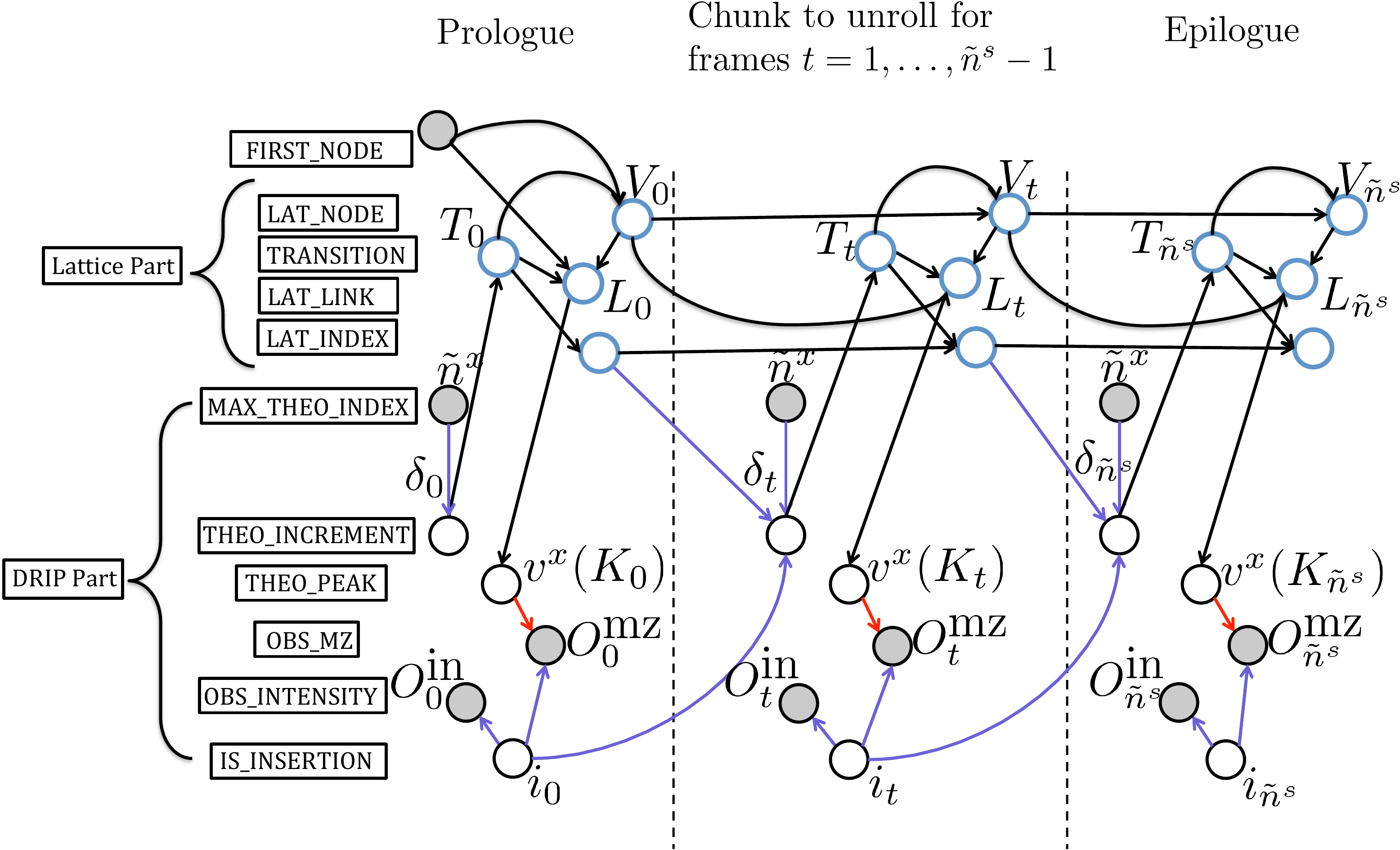}
\vspace{-0.1in}
\caption{Lattice DRIP Model}
\label{fig:latDRIP}

\end{minipage}
\vspace{-0.35in}
\end{figure}


Fig.~\ref{fig:latRep} shows the lattice representaion as dynamic graphical model structures.
$V_t = n_i$, $T_{t+1} = d$ ($d \geq 0$) determines the set of nodes
$V_{t+1}$ can take: $V_{t+1} = \{n_j \in N \mbox{ } | \exists p \in
P(G, n_i, n_j), |p| = d\}$.  $V_t = n_i$, $V_{t + 1} = n_j$, and $T_{t +1}
= d$, together determines the set of edges $L_{t + 1} = \{e \in E
\mbox{ } | p \in P(G, n_i, n_j), |p| = d, p[d - 1] = e\}$. In another
word, $L_{t + 1}$ is a random variable corresponding to all edges that
go into $n_j$ and can be reached from $n_i$ with a path of length
$d$. For the simple case, we may enforce $T_t \in \{0, 1\}$. If
$T_t = 0$, we stay at the current node, and if $T_{t+1} = 1$, $L_{t+1}$ becomes
the set of outgoing edges from $n_i$, and $V_{t+1}$ becomes the set of
destination nodes of the outgoing edges. ``FIRST\_NODE'' vertex has the observatoin
of the source node value $s$, and is used for initializing the time dependent structure.


The complexity for constructing the deterministic conditional probability table (CPT), which stores $Pr(L_{t} | V_t, V_{t + 1}, T_{t+1})$, for traversing the lattice as stated above is $|\{(e_1, e_2) \mbox{ }| e_1, e_2 \in E, \exists p \in P(G, e_1, e_2), |p| \leq d_{\max}\}|$, where $d_{\max}$ is the largest value $T_t$ can take, or in another word, maximum number of edges to ``jump over''. Please note that the CPT can be very sparse depending on the structure of the lattice as there may not exist a path between two nodes $n_i$ and $n_j$. The value of $d_{\max}$ can vary based on the underlying graphical model. If only edge-by-edge traversal of the lattice is required, than $d_{\max} = 1$. For DRIP, which we will talk in details later, $d_{\max} = \max_{x \in X} |x|$, which is the length of the longest candidate peptide. However, we may constrain $d_{\max}$ to take much smaller value than $\max_{x \in X} |x|$, as such long jumps are very unlikely in probability, so that we trade-off significant speed-ups with negligible performance loss.

Please note that the representation of lattices in GM can suit for any underlying dynamic graphical models, which access data in a streaming-like manner (do not go back and access previous data). Rather than accessing data in the traditional way, the lattice representation brings two major benefits: 1) various pruning strategies can be applied to speed-up the underlying GM significantly; 2) such compressed representation enables possibility of certain expensive learning methods, which requires to access the entire set of data, such as discriminative training.

\vspace{-0.1in}
\subsection{Lattice with DRIP}

Naturally, we convert the input of DRIP, which is a set of theoretical peptide peaks $v^x$ within certain mass window around the precursor mass, into a peptide peak lattice $G_p$. The alphabet $\Sigma_p$ for $G_p$ contains all possible peak m/z values (in Daltons) rounded to the nearest integer values. 
We build a lattice for each set of peptide candidates within certain mass window. The lattices then become the ``database'' to search for the best peptide candidate match. We don't count lattice construction time as an overhead as we can reuse the lattices for future queries just like a database.


Fig~.\ref{fig:latDRIP} shows the complete graphical model for Lattice DRIP. The ``THEO\_INCREMENT'' vertex $\delta_t$, which controls the number of deletions in DRIP, behaves similarly as the transition random variable $T_t$ in the lattice representation in GM, and we feed the value of $\delta_t$ into $T_t$. ``LAT\_INDEX'' keeps track of the number of edges from the current node to $s$. At first frame, the max value $T_0$ can take is then the max value of $\delta_t$, which is capped at $\max_{x \in X} |x|$. Over time frames, the max value of $T_t$ decreases as the rest lengths of peptides decrease. Different from the original DRIP model, where the rest length of each peptide candidate can be specified exactly, as only one peptide is process as a time, it is hard to map from the set of edges traversed so far to one specific peptide candidate, for the reason that multiple peptides can share those set of edges. As an approximation, we choose to use the rest length of the longest peptide encoded in the lattice as the rest lengths for all peptides. Theoretically, the Lattice DRIP score of a peptide would be a lower bound on the peptide's original DRIP score, as $Pr(\delta_t = a | \max(\delta_t) = b) < Pr(\delta_t = a | \max(\delta_t) = c)$ if $b > c > a$. In practice, we find such effect negligible, as we will show in the result session. 

The value of $L_t$, which contains the set of m/z values of b/y-ions, is fed into the ``THEO\_PEAK'' vertex $v^x(K_t)$ for scoring based on the mean Gaussians and the intensity Gaussian. The lattice acts like a ``database'' for querying theoretical peaks, and it does not affect the mechanism of the underlying GM. 

\vspace{-0.1in}
\subsection{Speeding Up Graphical Models with Lattices}

Lattices are compressed representations of the encoded data instances. By applying lattices in graphical models, the state space for accessing all the data instances is smaller compared to the original case where each data instance is accessed separately. An intuitive illustration is to consider about the ``simple lattice'', which contains $|X|$ disjoint paths from $s$ to $t$, and each path corresponds to one data instance. The state space for the simple lattice is no different from accessing each data instance separately. By constructing the lattice, common structures in the simple lattice get merged into shared edges so that the state space is smaller. Depending on the input data instances, the state space reduction can be significant. As data gets larger, we would expect more shared structures, so that the sizes of lattices grow only sublinearly. For the task of mass spectrometry, there are often hundreds of candidate peptides within a certain mass window for one spectrum, out of which around up to 50\% of peaks can be identified as redundant by the lattices. PTMs and other modifications to the peptides may enlarge the number of candidates to search tremendously, where lattices can perform more efficiently.

Beam pruning, which is a heuristic algorithm that expands only a pruned set of hypotheses in graphical model inference, fit perfectly with lattices. Different beam pruning strategies such as k-beam pruning, which preserves only k hypothesis states at each time frame $t$, shrinks the state space of graphical model inference significantly.
The hypothesis states of the original DRIP only consist of various deletion/insertion sequences of a single peptide candidate. Therefore, when various beam pruning strategies are applied, bad deletion/insertion sequences of the peptide get pruned away, yet we still end up evaluating every peptide candidate even though the best deletion/insertion sequences of certain peptides don't match the spectrum well. With lattices, beam pruning methods force peptide candidates filter themselves collaboratively. When applying beam pruning methods with lattices, the state space consists of various deletion/insertion sequences of all the peptide candidates. The badly scored peptides get pruned early on, and we end up scoring only a subset of the candidates. 


Reduce the maximum value of $T_t$ also decreases the search space of lattice. As stated above, complexity of the lattice traversing deterministic CPT is $|\{(e_1, e_2) \mbox{ }| e_1, e_2 \in E, \exists p \in P(G, e_1, e_2), |p| \leq d_{\max}\}|$. For DRIP, longer ``jumps'' or deletions are attributed with smaller probabilities. From Fig.~\ref{fig:jump}, which depicts the histogram of the length of deletions, long jumps are unlikely for the best-matching candidate. Therefore, the maximum value of $T_t$ can be set small, and rarely do we lose any good candidate. In practice, for Lattice DRIP we set the maximum allowed value of $T_t$ to 20.

There are enormous other ways to prune the lattices to accelerate the graphical model inference thanks to the lattices' capability of identifying common structures within the data. Pruning lattices statically like pruning trees, as well as embed probabilities into lattice edges to get pruning subject to certain distributions, may achieve good speed-ups, while the beam pruning and the limit of the longest jump, as we will mention in the result section, have gained us 7 to 15 times of acceleration.


\vspace{-0.1in}
\section{Training DRIP}\label{section:training}
\vspace{-0.1in}
DRIP is a highly trainable model amenable to
learning its Gaussian parameters, which provides
both a tool to explore the nonlinear m/z offsets caused by machine
error as well as a significant boost in performance.  For the 
overall training procedure, assume that we have a collection,
$\trainSet$, of $N$  i.i.d. pairs $(\bxt, \bzt)$, where $\bxt$ is an
observed spectrum and $\bzt$ the corresponding PSM we have strong
evidence to believe generated $\bxt$.  Let $\theta$ be the set of
parameters for which we would like to learn, in our case DRIP's
Gaussian parameters.  For generative training, we then wish to find
$\theta^* = \argmax _{\theta}\sum_{i=1}^N p(\bxt | \bzt, \theta)$,
i.e. we wish to maximize DRIP's likelihood with respect to the
parameters to be learned, achieved via expectation-maximization
(EM).

A much more difficult training paradigm is that of discriminative
training, wherein we do not simply wish to maximize the likelihood
under a set of parameters, but we would also like to simultaneously
minimize a parameterized distribution defined over a set of
alternative hypotheses.  In our case, this alternative set of
hypotheses consists of all candidate peptides within the precursor
mass tolerance not equal to $\bzt$, i.e. all incorrect explanations of
$\bxt$.  More  formally, our discriminative training criterion is that
of Maximum Mutual Information Estimation (MMIE); defining the set of
candidate peptides for $\bxt$ within precursor mass tolerance $w$ as
$\ctrain = \candidatePeps(\obsMz, \obsCharge, \pepDb, \mzTol)$ and
denoting the set of all training spectra and high-confidence PSMs as
$\sSet$ and $\xSet$, respectively, the function we would like to
maximize with respect to $\theta$ is then

\vspace{-0.1in}
\begin{eqnarray}
I_{\theta}(\sSet; \xSet) &=& \expect \log 
\frac{p(\bxt, \bzt|\theta)}{p(\bxt|\theta)p(\bzt|\theta)}
= \sum_{\bxt, \ctrain} p(\bxt, \ctrain |\theta)\log
\frac{p(\bxt, \bzt | \theta)}{p(\bxt|\theta)p(\bzt|\theta)}
= \sum_{\bxt, \ctrain} p(\bxt,\ctrain |\theta)\log
\frac{p(\bxt|\bzt,\theta)p(\bzt|\theta)}{p(\bxt |
  \theta)p(\bzt|\theta)} \nonumber \\
&= & \sum_{\bxt, \ctrain} p(\bxt, \ctrain |\theta)\log
\frac{p(\bxt|\bzt,\theta)}{p(\bxt | \theta)}  = \sum_{\bxt,\ctrain}
p(\bxt, \ctrain |\theta)\log \frac{p(\bxt|\bzt,\theta)}{\sum_{\pep \in
    \ctrain} p(\bxt, \pep |\theta)}.\label{eq:mmi}
\end{eqnarray}
\vspace{-0.1in}


We approximate this objective using the quantity $\tilde{I}_{\theta}(\sSet; \xSet) =
\frac{1}{N} \sum_{i = 1}^N \log
  \frac{p(\bxt |\bzt,\theta)}{\sum_{x \in \ctrain }p(\bxt,x |
    \theta)}$, which converges to the quantity in
  Equation~\ref{eq:mmi} for large $N$ by the i.i.d. assumption and the
  weak law of large numbers. Our objective is then
\vspace{-0.1in}
\begin{eqnarray}
\max_{\theta}\tilde{I}_{\theta}(\sSet; \xSet) &=&
\max_{\theta} \frac{1}{N} \sum_{i = 1}^N \log
\frac{p(\bxt |\bzt,\theta)}{\sum_{x \in \ctrain} p(\bxt, x | \theta)} = \max_{\theta} \frac{1}{N} \sum_{i = 1}^N \left ( \log
p(\bxt |\bzt,\theta) - \log \sum_{\pep \in \ctrain }p(\bxt, \pep | \theta) \right ),\label{eq:mmie}
\end{eqnarray}
where, for obvious reasons, we call $M_{n}(\bxt, \bzt) = \log p(\bxt |\bzt,\theta)$ the \emph{numerator model} and $M_{d}(\bzt) = \log \sum_{\pep \in \ctrain }p(\bxt,\pep | \theta)$ the \emph{denominator model}.  Note that the numerator model is our objective function for generative training, and in general the sum over possible peptide candidates in the denominator model makes the discriminative training objective difficult to compute.  However, by using lattices to efficiently perform the computation in the denominator model, we solve Equation~\ref{eq:mmie} using stochastic gradient ascent.

In stochastic gradient ascent, we calculate the gradient of the objective function with regard to a single training instance,
\vspace{-0.1in}
\begin{eqnarray}
\nabla_{\theta}\tilde{I}_{\theta}(\bxt; \bzt) &=&
\nabla_{\theta}M_n(\bxt | \bzt \theta) - \nabla_{\theta}M_d(\bxt | \theta),\label{eq:mmieGrad}
\end{eqnarray}
where the gradients of $M_n$ and $M_d$ are vectors referred to as \emph{Fisher scores}.  Correspondingly, we 
update the parameters $\theta$ using the previous parameters plus a damped version of Equation~\ref{eq:mmieGrad}, 
iterating this process until convergence.  The
overall algorithm is detailed in Algorithm~\ref{sga}.  In practice, we
begin the algorithm by initializing 
$\theta_0$ to a good initial value, i.e. the output of generative
training, and the learning rate $\eta_j$ 
is updated with $\eta_{j+1} = (\sqrt{j})^{-1}$.
Intuitively, the gradients move in the direction maximizing the
difference between the numerator and denominator models,
encouraging improvement for the numerator while discriminating against
the incorrect labels in the denominator.  Our experimental results
show that discriminative training positively influences performance (Section~\ref{section:discTrainingResults}).

\vspace{-0.2in}
\begin{algorithm}
\caption{Discriminative Training via Stochastic Gradient Ascent}\label{sga}
\begin{algorithmic}[1]

\State Initialize $\theta_0$, $\eta_0$. Let $j = 0$.

\While{True}
\State $\theta_{j + 1} := \theta_{j}$; $j:= j+1$;
\State Update learning rate $\eta_j$;

\For{$(\bxt ,\bzt ) \in \trainSet $}
	\State  $\theta_{j} := \theta_j + \eta_{j}
	(\nabla_{\theta}M_n(\bxt) | \bzt \theta) -
      \nabla_{\theta}M_d(\bxt | \theta))$; \label{algorithm:stochasticUpdate}

\EndFor

\If{$||\frac{\theta_{j} - \theta_{j - 1} }{\theta_{j - 1}}||  < \epsilon$}
	\State break;
\EndIf

\EndWhile
\end{algorithmic}
\end{algorithm}
\vspace{-0.4in}

\subsection{Discriminative Training with Lattices}
\vspace{-5pt}
Discriminative training is expensive to execute. The denominator model
requires calculating the gradients of all possible candidate peptides
$\ctrain$, which can be infeasible for many tasks, yet to represent
all possible labels in some graphical models, e.g. DRIP, is neither a
trivial task. The hardness of representing all possible labels in DRIP
comes from the fact that it is difficult to constrain the model to consider
valid peptides only, as the distance between subsequent theoretical
peaks can take on any value.
\JoTS{maybe explicitly say it would be
  nontrivial to construct a graphical model with hidden variables in
  order to sequence through all possible valid theoretical spectra of
  peptides in $\ctrain$, since valid units between theoretical peaks
  may be almost any arbitrary value; this is juxtaposed with Didea
  where it is a simple matter to make the amino acids hidden and make
  sure the cterm mass is observed to the precursor mass in the
  prologue and the nterm mass is observed to the precursor mass in the
  epilogue.}

Lattices work perfectly with discriminative training in solving the scalability and representability problem. The lattice of all possible labels can be very compact, together with different strategies to speed up graphical model with lattices discussed in the previous session, discriminative training with lattice can be quite feasible  and efficient. 

The Lattice is a general framework to represent hypotheses for any dynamic graphical model. The denominator model of the discriminative training for DRIP is exactly the same as the Lattice DRIP model. Even for graphical models, which are capable of encoding all possible labels for discriminative training, lattices have the advantage of being more efficient and amenable to various modifications, as we can encode any set of labels into the lattice, so that discriminative training against any distribution is achievable.

\vspace{-0.2in}
\section{Experimental methods}
\vspace{-0.1in}
We benchmark all methods using three data sets: one low-resolution
data set from the yeast {\em S.\ cerevisiae} consisting of
35,236 spectra (Yeast), one low-resolution data set from {\em C.\
  elegans} consisting of 23,697 spectra (Worm-I), and one
high-resolution {\em C.\ elegans} dataset consisting of 7,557
spectra (Worm-II).  Further details regarding the Yeast and Worm-I
datasets and corresponding target databases may be found
in~\cite{kall:semi-supervised}, and details regarding Worm-II and
its database may be found in~\cite{howbert:computing}.  The datasets
and target databases are available on the corresponding supplementary
pages.

In order to ensure that all methods score exactly the same peptides,
each search engine was provided with a pre-digested set of peptide
sequences, rather than intact proteins sequences.  Each protein
database was digested using trypsin without suppression of cleavage by
proline.  Precursor charges range from 1+ to 3+ for Worm-I
and Yeast and from 1+ to 5+ for Worm-II.  For spectra with
multiple charge states, the top scoring PSM was chosen per method.

All search algorithms were run with as equivalent settings as
possible; machines were specified to CID, no allowance for missed cleavages
 or isotope errors, and a single fixed
carbamidomethyl modification.  For the two data sets with
low-resolution precursors (Yeast and Worm-I), the mass tolerance $w$
was set to $\pm$ 3~Th, whereas for Worm-II $w$ was set to $\pm$ 10~ppm.
Comet searches (XCorr E-values) used flanking peaks around each b- and
y-ion, whereas Tide searches (XCorr and XCorr p-values) did not.
These settings provided optimal performance for the two methods.  All
benchmark methods included neutral loss peaks in theoretical spectra,
although these peaks are not modeled in DRIP.

A significant challenge in evaluating the quality of a spectrum
identification algorithm is the absence of a ``ground truth'' data set
where the generating spectrum is known for each observed spectrum.  We
therefore follow the standard approach of using {\em decoy peptides}
(which in our case correspond to randomly shuffled versions of each
real {\em target} peptide) to estimate the number of incorrect
identifications in a given set of identified spectra.  In this work,
targets and decoys are scored separately and used to estimate the
number of identified matches at a particular \emph{false discovery
  rate} (FDR), i.e. the fraction of spectra improperly identified at a
given significance threshold \cite{kall:assigning}.  Because the FDR
is not monotonic with respect to the underlying score, we instead use
the \emph{q-value}, defined to be the minimum FDR threshold at which a
given score is deemed to be significant.  Since data sets containing
more than 10\% incorrect identifications are generally not practically
useful, we only plot $q$-values in the range $[0,0.1]$.

\vspace{-0.2in}
\section{Results}
\vspace{-0.1in}
\subsection{Charge calibration improves statistical power}
\begin{figure*}
\vspace{-29pt}
  \centering
  \subfigure[Worm-I]{\label{fig:worm}\includegraphics[width=0.27\linewidth]{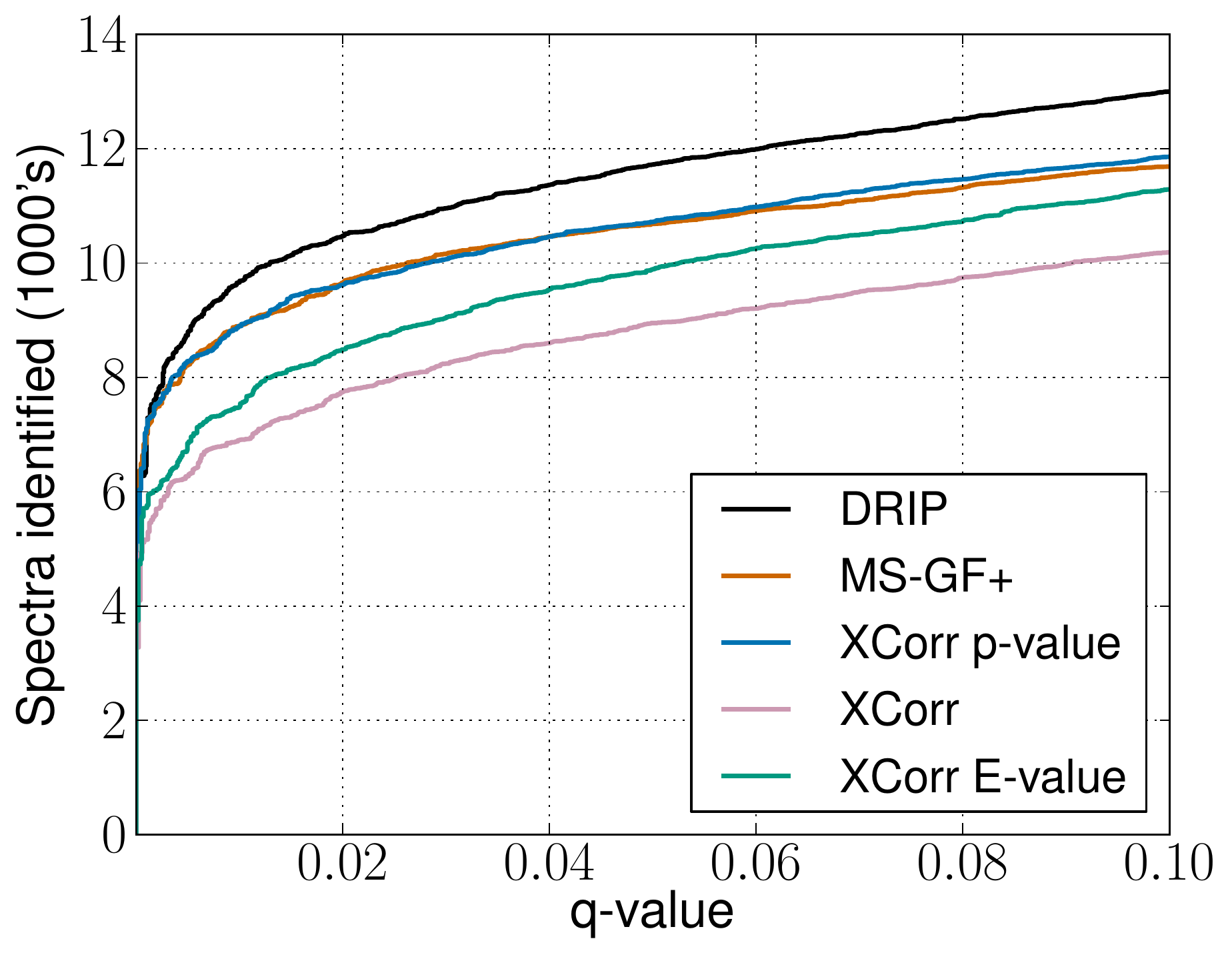}}
  \subfigure[Yeast]{\label{fig:yeast}\includegraphics[width=0.27\linewidth]{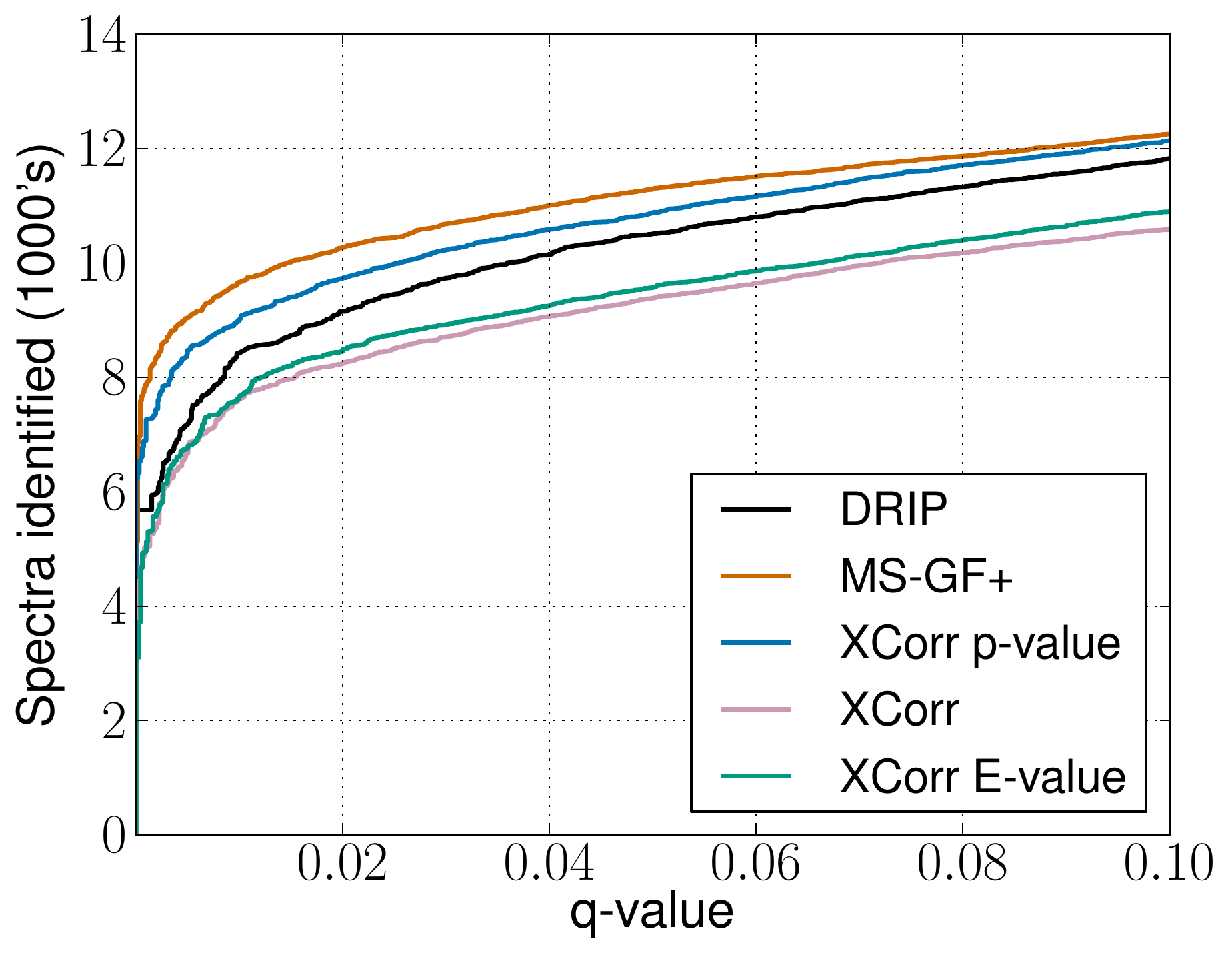}}
  \subfigure[Worm-II]{\label{fig:wormHighRes}\includegraphics[width=0.27\linewidth]{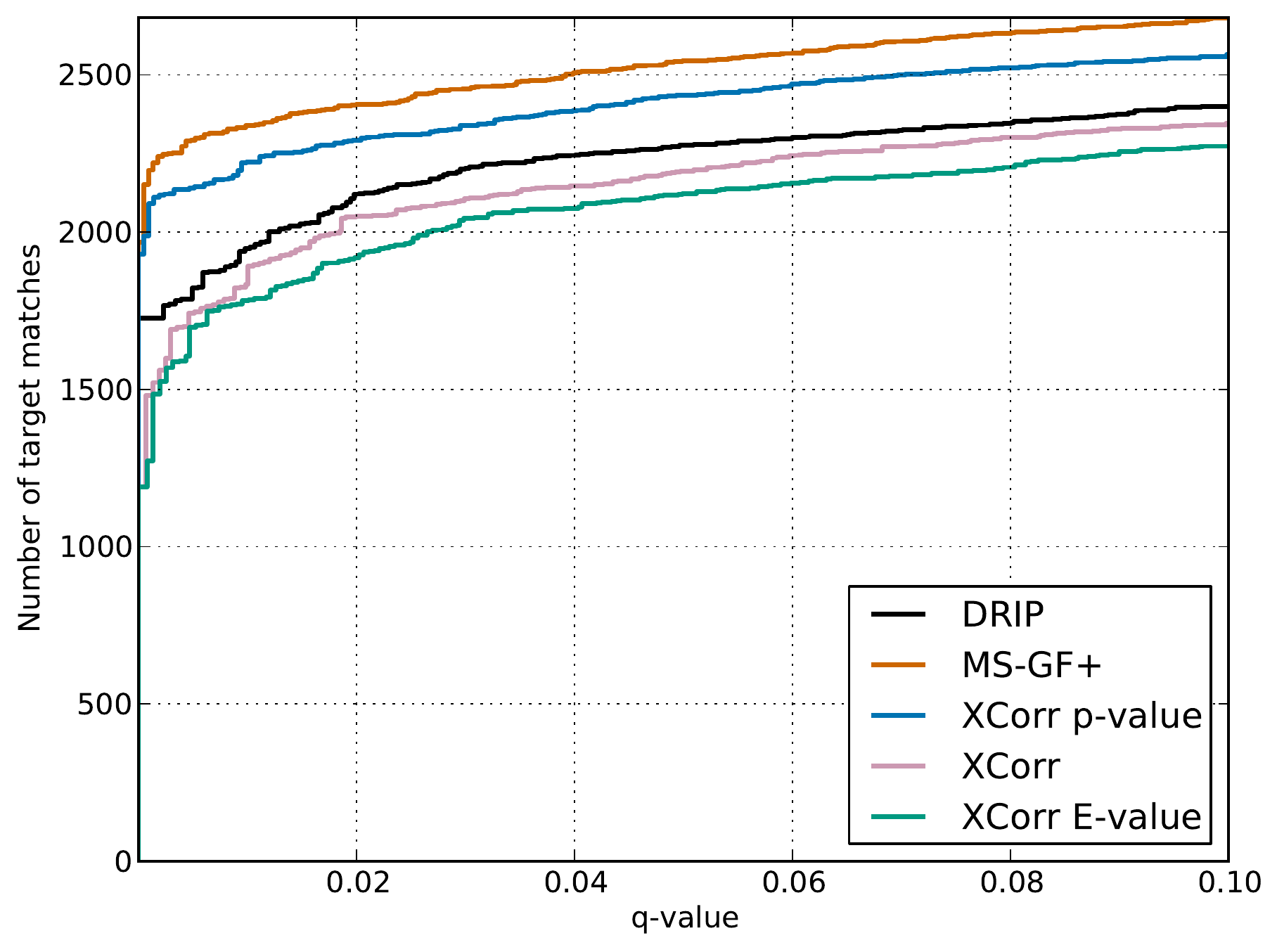}}
\vspace{-0.2in}
  \caption{Charge variation results for all datasets.}
  \label{fig:chargeVaryingAbsRanking}
\vspace{-17pt}
\end{figure*}

DRIP charge
recalibrated outperforms XCorr and XCorr E-value on all
datasets, and outperforms
all datasets on Worm-I by a large margin.  Though DRIP is outperformed
by the two p-value methods on Yeast-I and
Worm-II, this gap may be closed by gathering training data
better representative of the corresponding spectra and their charge
states (the high quality training data used only contains
charge 2+ PSMs), which is currently underway.  Furthermore, with the
incorporation of lattices into DRIP, we now have a mechanism
by which to sequence through entire sets of peptides.  We believe this
to be a critical step towards evaluating DRIP score thresholds with
respect to arbitrary peptide sets, thus allowing the computation of DRIP
p-values for which a performance increase such as XCorr to XCorr p-value
is expected.  It is worth noting that while MS-GF+ and XCorr p-value
evaluate the entire set of possible peptides equal to a given
precursor mass, the utilization of lattices allows DRIP to evaluate an
arbitrary set of peptides (even regardless of precursor mass), which
is strictly more general and flexible.

\vspace{-0.15in}
\subsection{Faster DRIP scoring using lattices}
\vspace{-0.1in}

\begin{figure}
\vspace{-0.25in}
\hspace{-0.2in}
  \begin{minipage}[b]{0.5\linewidth}

\centering
\includegraphics[page=1,trim=0.1in 0.05in 0.3in 0.1in,clip=true,
width=0.45\linewidth]{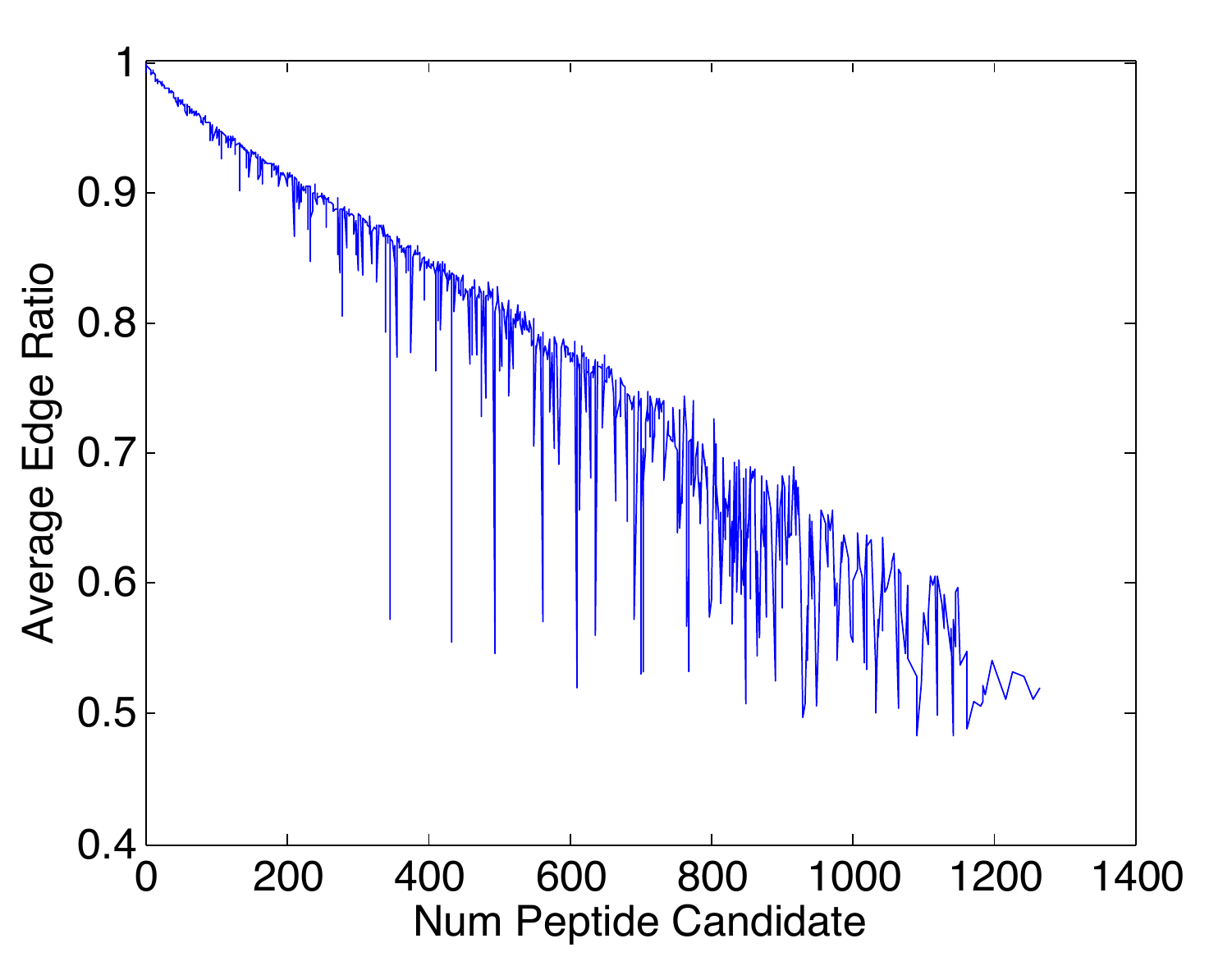}
\includegraphics[page=1,trim=0.1in 0.05in 0.3in 0.1in,clip=true,
width=0.45\linewidth]{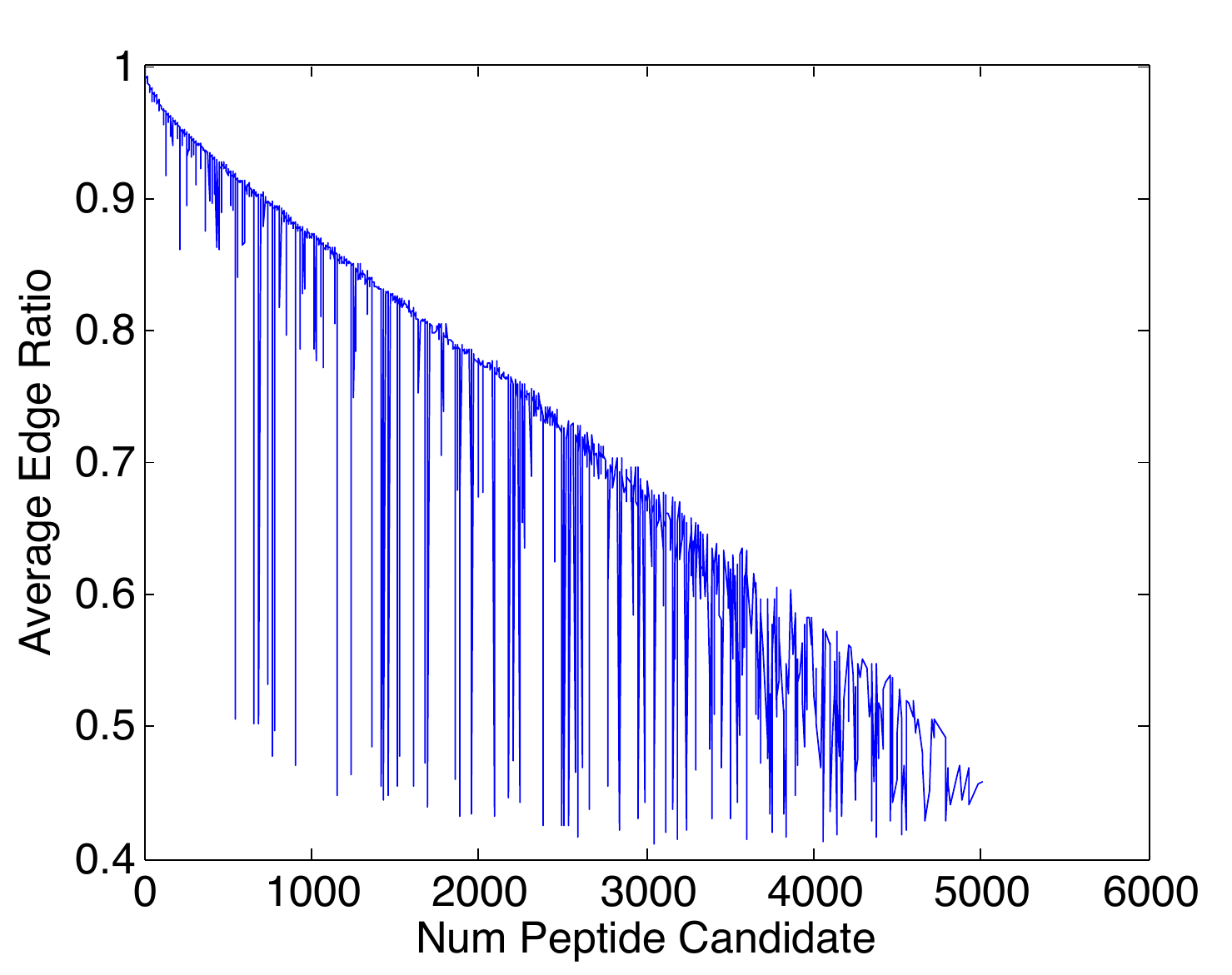}
\vspace{-0.1in}
\caption{Lattice Compression Ratio: for mass windows of certain number of peptide candidates,
calculate the number of edges in the corresponding lattices divided by the number of peaks
in the candidates from Yeast(left)/Worm(right) databases.
}
\label{fig:latEdgeCount}

  \end{minipage}%
\hspace{0.1in}
  \begin{minipage}[b]{0.5\linewidth}

\includegraphics[page=1,trim=0.0in 0.1in 0.0in 0.1in,clip=true,
width=0.5\linewidth]{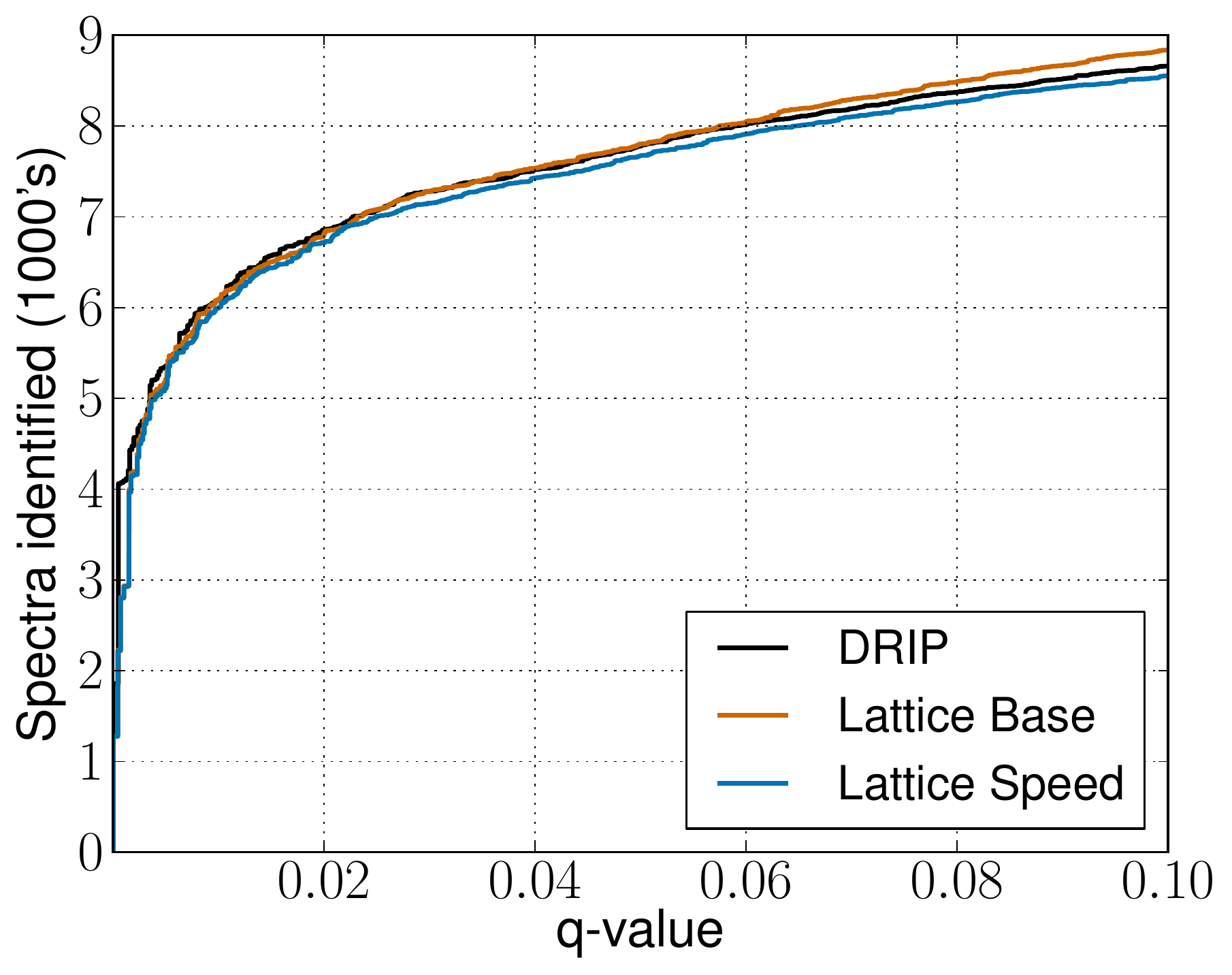}
\includegraphics[page=1,trim=0.0in 0.1in 0.0in 0.1in,clip=true,
width=0.5\linewidth]{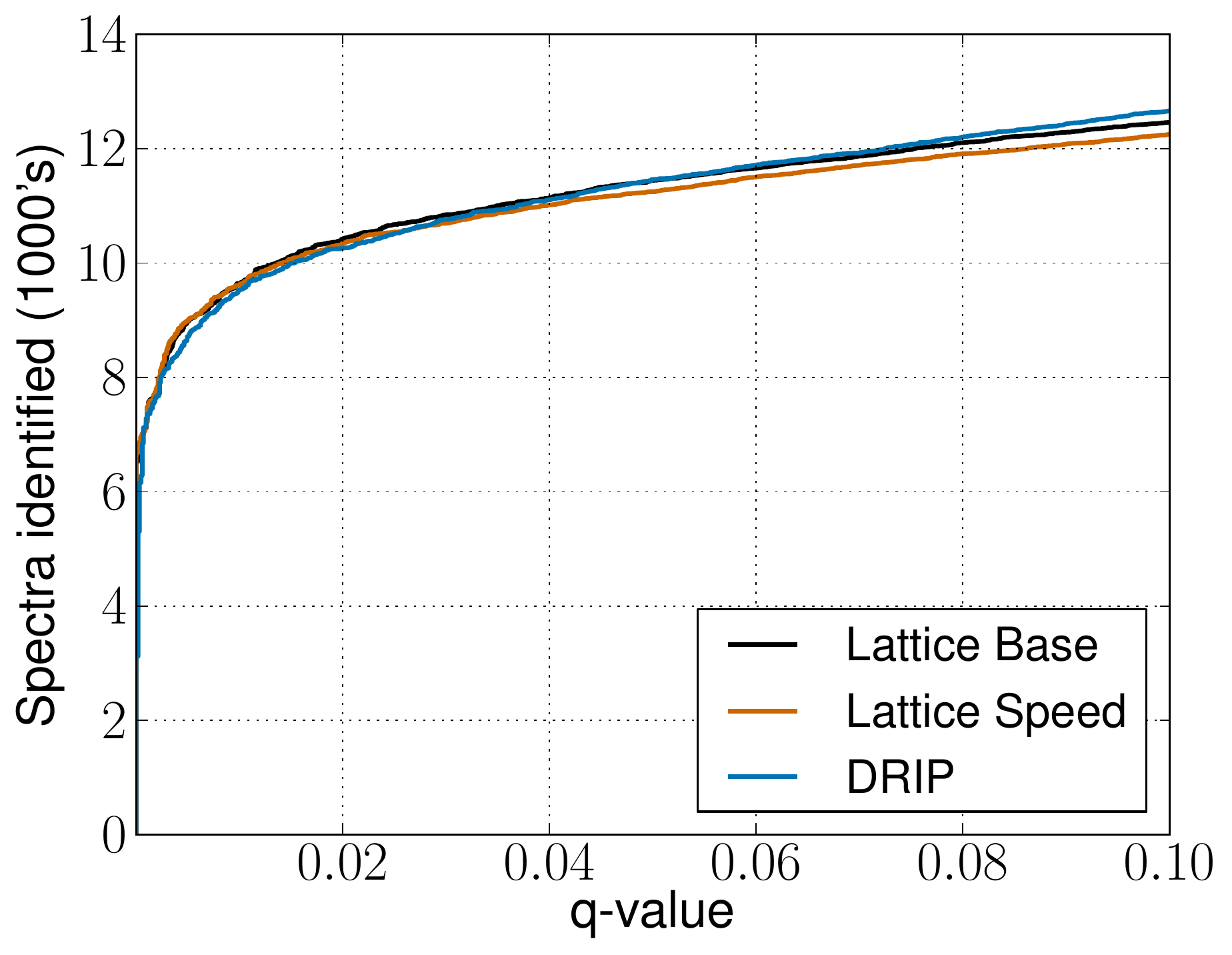}
\caption{Performance curves: compare two lattice results
  against the original DRIP, on Yeast (left) and Worm-I (right)
  with charge 2.}
\label{fig:latCurve}
\end{minipage}
\end{figure}

We first present results on the ability of lattices to compress
peptides.  We build lattices on four peptide datasets with theoretical
peak values as the alphabet. Figure~\ref{fig:latEdgeCount} shows
the effectiveness of compression as the number of peptide candidates
in the precursor mass window (size of $\pm$ 3 \thomson) varies. The
compression ratio is defined to be the ratio of the number of edges in
the lattices, which are built on mass windows of certain number of peptide cadidates,
to the number of theoretical peaks in the candidates in those windows. We
note that the compression ratio decreases almost linearly as the
number of peptides increases, and such reduction can be over $50\%$.


To illustrate the effectiveness of improving inference time using
lattices, we test Lattice DRIP with the following beam pruning
strategies, and compare the results against DRIP (using the beam
pruning settings described in \cite{halloran2014uai-drip}):
\begin{itemize}
\item
$lattice_{base}$: pruning with $k$-beams which are dynamic across time frames, with wider beams for the early part and narrower beams later on.
\item
$lattice_{speed}$: pruning aggressively with dynamic $k$-beam (same as
$lattice_{base}$) with narrower beams. Moreover, the pbeam pruning
algorithm is applied, which prunes the state space while building up
the inference structures.
\end{itemize}


\begin{wraptable}{r}{0.35\textwidth}
\caption{Percentage running time of lattice methods compared to original DRIP.}\label{tab:timing}
\vspace{-10pt}
  \begin{tabular}{| l || c | c | r |}
    \hline
    dataset & $lattice_{speed}$ & $lattice_{base}$ & DRIP \\ \hline
    Yeast & 7.78\% & 14.80\% & 100\%\\ \hline
    Worm-I & 8.59\% & 16.36\% & 100\%\\ \hline
  \end{tabular}
\vspace{-20pt}
\end{wraptable}
We note that while the space of beam pruning strategies is large and
it is possible that the two above methods are not optimal in terms of
absolute computational efficiency achievable using lattices, they give
significant speed improvements over the original DRIP.
To evaluate the speed-ups possible with lattices, we randomly select
50 spectra from Yeast and Worm-I, respectively, and record inference
time. We use the same graphical model inference engine for
all three methods. Experiments were carried out on a 3.40GHz CPU with
16G memory. The lowest CPU time out of 3 runs is recorded, and we
report the relative CPU time of lattice methods to original DRIP in
Table~\ref{tab:timing}.  Timing tests show that utilizing lattices,
DRIP runs 7-15 times faster.  Note that this comes at practically
no expense to performance, as show in Figure~\ref{fig:latCurve}.

\subsection{Discriminative training further boosts statistical power}\label{section:discTrainingResults}
\begin{figure}[htbp!]
\vspace{-15pt}
\centering
\subfigure[Yeast, charge 2+]{\includegraphics[page=1,trim=0.0in 0.1in 0.0in 0.1in,clip=true,width=0.26\linewidth]{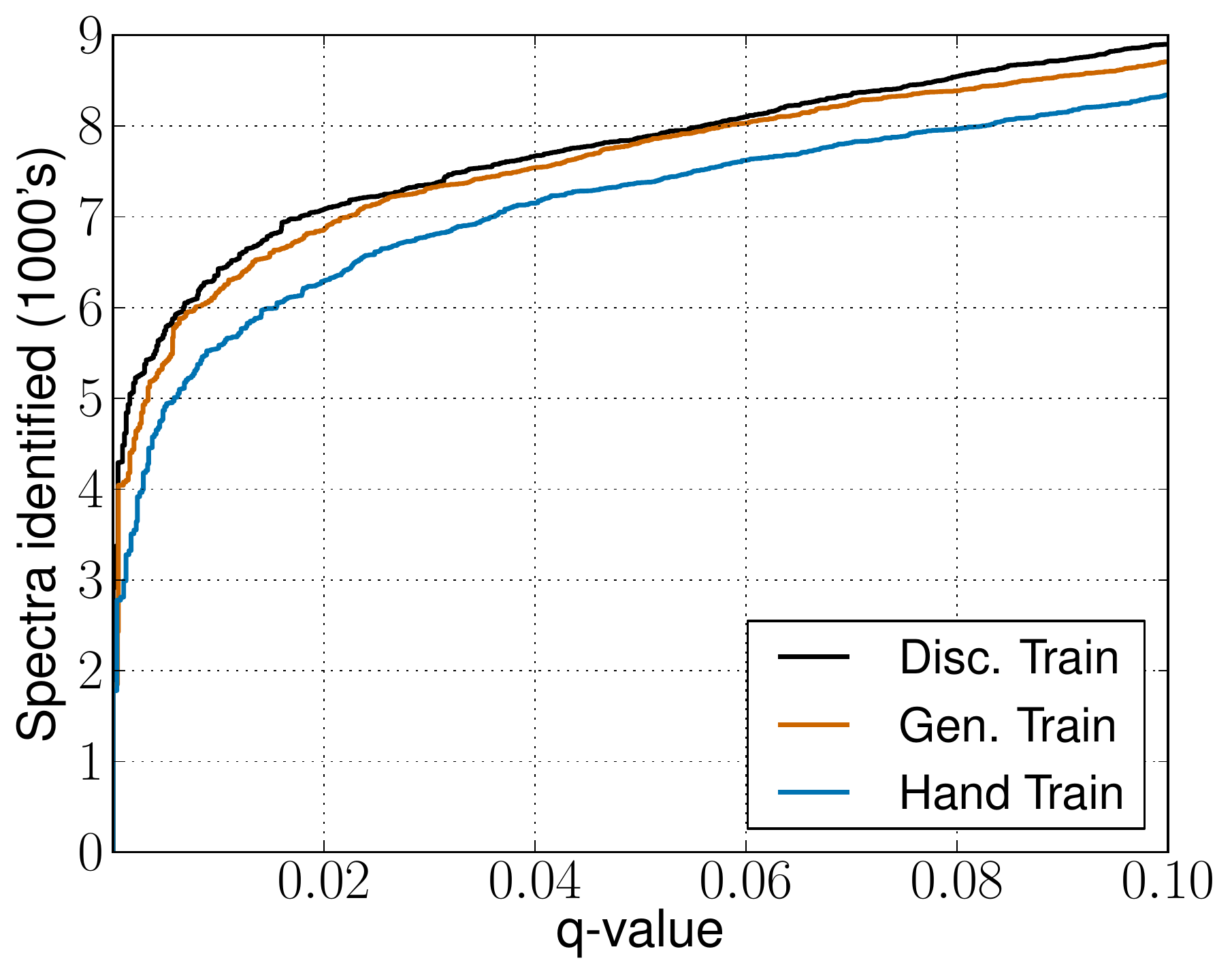}\label{fig:yeastDisc}}
\subfigure[Worm-I, charge 2+]{\includegraphics[page=1,trim=0.0in 0.1in 0.0in 0.1in,clip=true,width=0.26\linewidth]{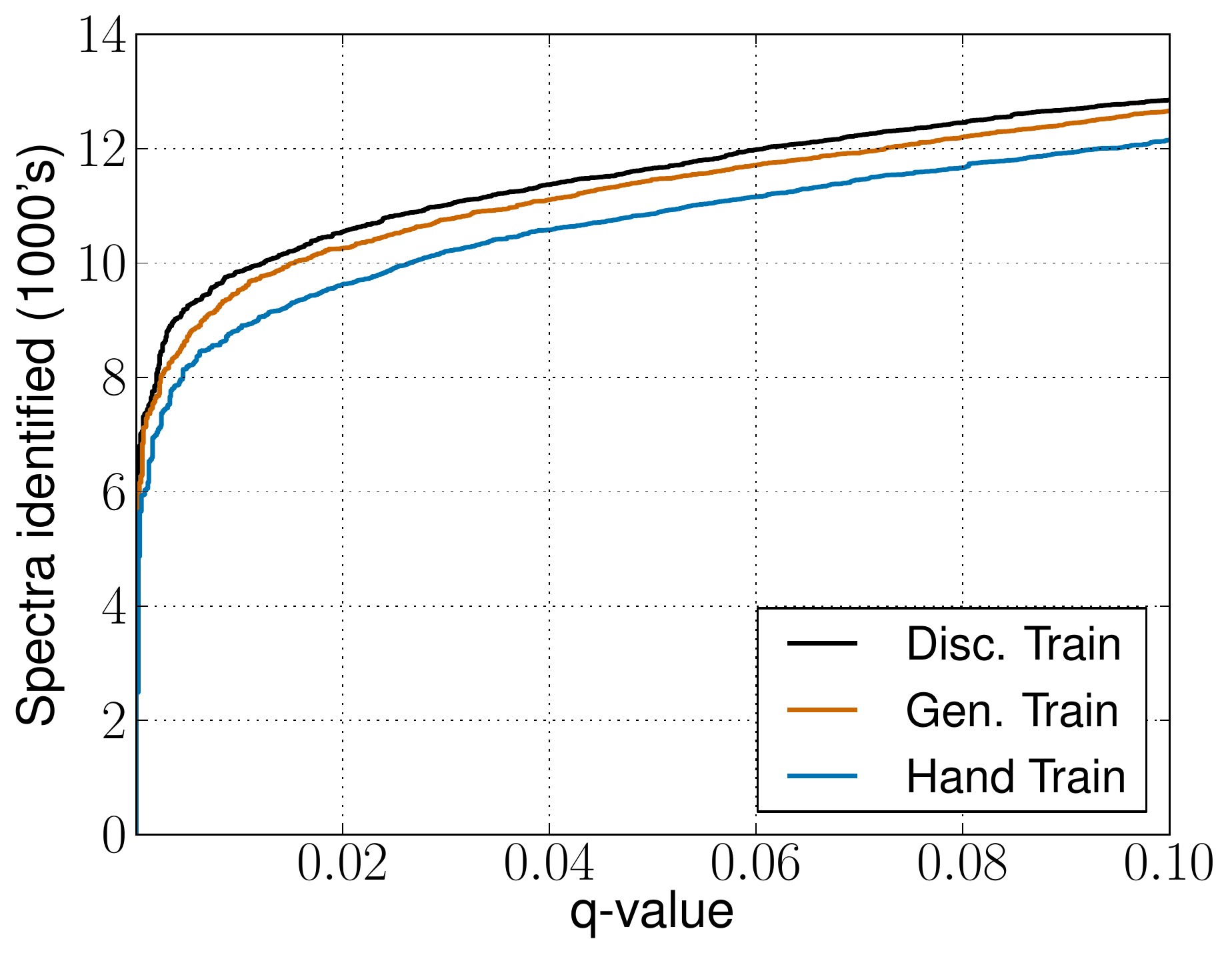}\label{fig:wormDisc}}
\subfigure[Means used to score observed
peaks.]{\includegraphics[page=1,trim=0.0in 0.1in 0.0in
  0.1in,clip=true,width=0.36	\linewidth]{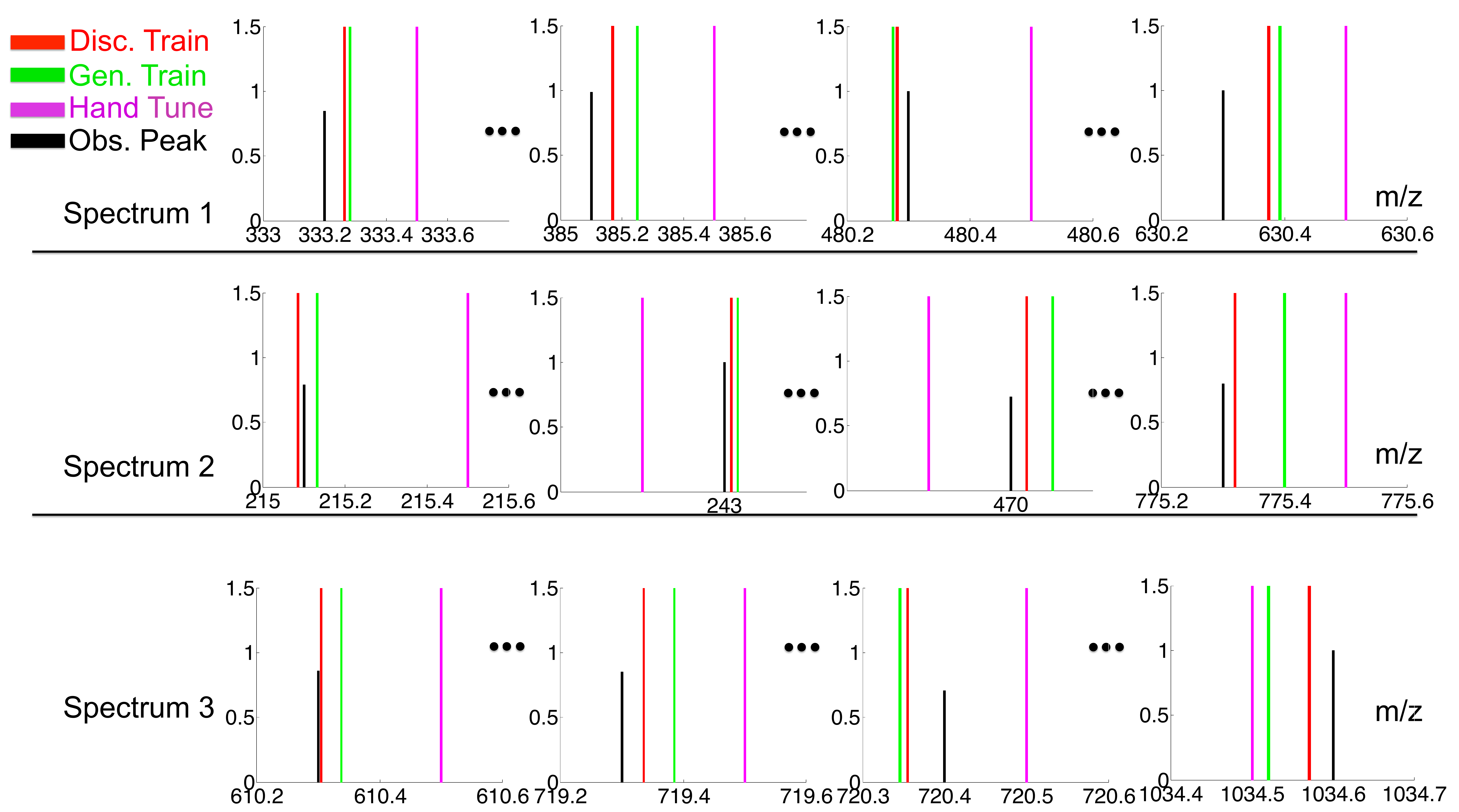} \label{fig:matchedMeans}}
\vspace{-0.2in}
\caption{Effects of training DRIP parameters.}
\label{fig:PeakMatch}
\vspace{-17pt}
\end{figure}

As detailed in Section~\ref{section:training}, we use a set of
high-confidence, charge 2+ PSMs and their corresponding peptide database
to discriminatively train DRIP, utilizing lattices in the denominator
model.  To illustrate the power afforded by DRIP's learning
capabilities, we also illustrate performance under hand-tuned
parameters, where DRIP's Gaussian means are placed between unit
intervals along the m/z access (representative of fixed
binning strategies).
Figures~\ref{fig:yeastDisc},~\ref{fig:wormDisc} show that the discriminatively
trained DRIP improves performance, especially for low $q$-values,
arguably the most important region of performance.  Note that the
discriminatively trained model depicted employs the $lattice_{base}$
pruning strategy, and thus we have an increase in accuracy as well an
approximately seven-fold speed-up.


In Figure~\ref{fig:matchedMeans}, we further investigate the influence of
training methods on the m/z Gaussian means of DRIP and their
performance. Choosing three spectra from Yeast at random, four high
intensities are displayed per spectrum, along with the means
used to score these peaks.  With the observed peaks plotted in black,
 the resulting discriminatively trained Gaussian means are much
closer than all other means, yielding better scoring and more accurate results.

\vspace{-0.2in}
\section{Conclusions}
\vspace{-0.1in}
In this paper, we show several significant improvements to the
DRIP model. We show how to apply word lattices to compress
peptide candidate sets in order to dramatically speed up inference in
DRIP (7-15 fold speed up).  With the ability to compactly represent
entire sets of peptides, we extend DRIP's learning framework to
discriminative training, leading to performance gains at low
$q$-values.  
We have also greatly simplified the DRIP model itself and allowed the
ability to accurately search varying charge states per dataset.

There are several avenues for future work.  With the ability to
effienctly sequence through entire set peptides afforded by lattices,
we will investigate ways to take thresholds with respect to DRIP
scores in order to compute p-values.  As evidenced by other scores for
which exact p-values are computed (MS-GF+, XCorr
p-value), this is expected to greatly increase DRIP's performance.  We
have also only scratched the surface of training parameters with the
model.  Collecting an assortment of training data, we will
discriminatively train the model for a host of different charge
states, machine types, and organisms in an effort to further increase
accuracy over the wide array of tandem mass spectra encountered in
practice.  We also plan to explore ways to alleviate the labor
intensive process of collecting high-confidence training PSMs
described in \cite{Klammer2008}, in order to simplify the overall
training process, beginning from data collection.


\clearpage
\bibliographystyle{splncs03}
\bibliography{refs,main}
\end{document}